\documentclass[twocolumn,superscriptaddress]{revtex4-1}
\usepackage{amsmath,amssymb}
\usepackage{graphicx,subfigure,epsfig}
\usepackage{hyperref}
\usepackage{geometry}
\usepackage{tablefootnote}

 \usepackage{booktabs}
 \usepackage{multirow}
 \usepackage[normalem]{ulem}
 \useunder{\uline}{\ul}{}

\hypersetup{
    colorlinks=true,
    citecolor=blue,
    linkcolor=red,
    filecolor=magenta,
    urlcolor=cyan,}

\begin{document}

\title{Probing hairy black holes caused by gravitational decoupling using quasinormal modes and greybody bounds}

\author{Yi Yang}
\email{yangyigz@yeah.net }
\affiliation{College of Physics, Guizhou University, Guiyang, 550025, China}

\author{Dong Liu}
\email{dongliuvv@yeah.net}
\affiliation{College of Physics, Guizhou University, Guiyang, 550025, China}

\author{Ali \"Ovg\"un}
\email{ali.ovgun@emu.edu.tr}
\affiliation{Physics Department, Eastern Mediterranean University, Famagusta, 99628 North Cyprus via Mersin 10, Turkey.}

\author{Zheng-Wen Long}
\email{zwlong@gzu.edu.cn (Corresponding author)}
\affiliation{College of Physics, Guizhou University, Guiyang, 550025, China}

\author{Zhaoyi Xu}
\email{zyxu@gzu.edu.cn}
\affiliation{College of Physics, Guizhou University, Guiyang, 550025, China}

\vspace{0.5cm}
\date{\today}

\begin{abstract}
Gravitational decoupling can add hair to the black holes by adding extra sources. The quasinormal modes of hairy black hole caused by gravitational decoupling for the massless scalar field, electromagnetic field, and gravitational perturbation are investigated. The equations of effective potential for three perturbations are derived in hairy black hole spacetime. We also study the time evolution corresponding to the three perturbations, and the quasinormal mode frequencies are calculated using the Prony method through the time-domain profiles. By analyzing the influence of the hairs ($\alpha$, $l_0$ and $Q$) for the black holes we studying on quasinormal mode, we find that the hairs $\alpha$ and $l_0$ decrease the oscillation frequency of the gravitational wave signal, and the hair $Q$ increase its oscillation frequency.Furthermore, we have calculated the bounds of greybody factor and high-energy absorption cross section with the Sinc approximation, which reveals that the presence of charges ($\alpha$ and $l_0$) generating primary hair can increase the probability of gravitational radiation arriving spatial infinity, whereas the charge $Q$ from the extra sources does the opposite.

\end{abstract}

%\pacs{95.30.Sf, 04.70.-s, 97.60.Lf, 04.50.+h}
%\keywords{Gravitational decoupling; Hairy black hole; Quasinormal modes; Greybody factor}
\maketitle
%\tableofcontents
%\newpage
%-----------------------------------------------------------------------------

\section{Introduction}
\label{sec:intro}
General relativity shows that when a massive stellar collapses into a black hole, there are only three physical quantities: mass, angular momentum, and electric charge, which uniquely determine the properties of the black hole. All the other information (``hair") disappeared \cite{nh1,nh2}. However, there may be some other physical quantities that describe black holes. For example, black hole may also have quantum hairs \cite{qh}. Many methods are used to evade the no-hair theorem \cite{Grumiller:2019fmp,evade1,evade2,evade3,evade4}.
In particular, Ovalle proposed a gravitational decoupling method to obtain the solutions of Einstein field equations by decoupling of the gravitational sources \cite{ovalle5,Ovalle:2018gic}. The gravitational decoupling method has gained a lot of attention \cite{Sharif:2018toc,Ovalle:2019lbs,Ovalle:2018ans,Ovalle:2017wqi,Rincon:2019jal,Gabbanelli:2018bhs,Contreras:2019iwm,Panotopoulos:2018law,Tello-Ortiz:2020ydf,ovalle1,ovalle2,ovalle3,ovalle4} mainly, because it has the following advantages: (i) it can decouple the complex energy-momentum tensor into relatively simpler components; (ii) one can use it to extend some known seed solutions to more complex solutions; (iii) one can use it to find solutions of gravitational theories other than general relativity (GR).
Ovalle et al. assume that there are additional general sources described by the conserved energy-momentum tensor  $\theta_{\mu \nu}$. This $\theta_{\mu \nu}$ can explain one or more fundamental fields, and its key property is that it is subject to gravity but does not directly interact with the matter of the black hole. Then, they obtained the hairy black hole solution by gravitational decoupling \cite{ovalle6,ovalle7}.
In particular, two new families of hairy black holes have been discovered by Ovalle et al. in Ref. \cite{ovalle7}, which demonstrates that basic deformations of the seeded Schwarzschild vacuum retain the energy conditions, and proposes a novel method for evading the no-hair theorem depending on a primary hair correlated with the charge that generates these transformations. This leads us to be interested in the stability of such hairy black hole.

The stability of a black hole under certain perturbation is closely related to the properties of the black hole itself. Usually, the stability of black hole spacetime under perturbation is studied through the field evolution of the black hole background or the black hole merger.
A perturbed black hole can emit gravitational wave (GW) which is dominated by quasinormal modes (QNM) \cite{Vishveshwara:1970zz} with complex frequency, where the real part of QNM frequency denotes the oscillation frequency of the black hole when perturbed, and the imaginary part represents the decay rate \cite{Kokkotas:1999bd,Nollert:1999ji,Cardoso:2019rvt,Konoplya:2011qq,Chirenti:2017mwe,Pantig:2022gih}. In addition to characterizing the stability of black hole spacetime, QNM frequency also plays a very important role in determining black hole parameters. Moreover, a collision event between black holes will go through three phases: inspiral, merger and ringdown phases. LIGO/VIRGO first observed gravitational waves produced from the black holes merge in 2016 \cite{LIGOScientific:2016aoc}. They find ringdown in the gravitational wave signal, which appears at the end of the waveform and consists of rapidly decaying oscillations. The ringdown phases is the QNM of the remnant black hole, which makes people generate great interest in the study of black hole QNM.
In Ref. \cite{Cheung:2021bol}, Cheung et al. studied the QNM spectrum in which the Schwarzschild potential is perturbed by a small ``bump'' consisting of the Gaussian potential or the P\"{o}schl-Teller potential, and they demonstrate that the fundamental mode is unstable under general perturbations.
In Ref. \cite{Boyanov:2022ark}, the authors studied the pseudospectrum of horizonless exotic compact objects (ECO) with reflective surfaces close to the Schwarzschild radius, and demonstrate that the QNMs of ECO are affected by the overall spectral instability.
In Ref. \cite{Chatzifotis:2021pak}, the authors studied the stability of a black hole with scalar hair under axial gravitational perturbations, and they find that this black hole with scalar hair is linearly stable.
Particularly, Cardoso et al. point out that the QNM spectra of some compact objects with the light rings are radically different from that of the black hole, but they still show a similar ringdown stage \cite{Cardoso:2016rao}. In fact, this especial QNM spectrum is also known as echoes \cite{Cardoso:2016oxy,Cardoso:2017cqb,Maggio:2019zyv,2022zym,Konoplya:2018yrp,Bronnikov:2019sbx,Churilova:2021tgn,Yang:2021cvh,Ou:2021efv,Vlachos:2021weq,LongoMicchi:2020cwm,Chowdhury:2020rfj,guo:2022umh,DuttaRoy:2019hij}.
Numerous studies have shown that the QNM frequency of a black hole is only determined by the properties of the black hole itself and fields \cite{Ferrari:2007dd,Berti:2009kk,Berti:2005ys,lijin2,lijin3,liudj,csb,csb2,Daghigh:2008jz,Daghigh:2011ty,Daghigh:2020mog,Zhidenko:2003wq,Zhidenko:2005mv,Chabab:2017knz,Lepe:2004kv,Gonzalez:2017shu,Lin:2016sch,Berti:2009kk,Konoplya:2003ii,Liu:2021xfb,Chakraborty:2017qve,Mishra:2021waw}. Therefore, we believe that the properties of hairy black hole with additional sources described by a conserved energy-momentum tensor can be inspected through the QNM.

Quantum mechanics and general relativity form the bedrock of the current understanding of physics, but the two theories don't to work together. In 1974, Hawking had showed that black holes are not perfectly ``black" but actually emit particles \cite{Hawking:1974rv,Hawking:1975vcx} as well as scatter, absorb and radiation. Hawking radiation propagates on the curved spacetime due to the black hole, and the curvature of spacetime behaves like a gravitational potential where the radiation is scattered from it. There are two parts: (i) Reflected back into black hole and (ii) transmitted to spatial infinity. Hence one can calculate the transmission probability, known as the greybody factor, using various methods such as the matching technique \cite{Fernando:2004ay},  the WKB approximation for the high gravitational potential \cite{Parikh:1999mf,hwkb1} and using the rigorous bound \cite{Boonserm:2008zg,Boonserm:2019mon,Visser:1998ke}. On the other hand, the evaporation rate is proportional to the total absorption cross section. Since Hawking's discovery, there have been many studies done on the absorption cross section of scalar field in black hole spacetime. First Sanchez \cite{Sanchez:1977si} had calculated the absorption cross section of massless scalar wave for the Schwarzschild black hole in the high-frequency regime, which exhibits oscillation around the geometrical optics limit using numerical methods. It is known that the low frequency behavior of the scalar field absorption by black hole tends to surface area of the event horizon \cite{Das:1996we}. At high energy, the absorption cross section oscillates around a limiting constant value. For a scalar field absorbed of a spherically symmetric black hole, using complex angular momentum methods, Decanini et al. had derived the Sanchez's result in the high-frequency regime with a more accurate coefficient. They obtain the absorption cross section using sinc-function $\sigma_{a b s}=-8 \pi e^{-\pi} \sigma_{\text {geo }} \operatorname{sinc}[2 \pi(3 \sqrt{3} M) \omega]$, with $\operatorname{sinc}(x)=(\sin x) / x$ and $\sigma_{g e o}=27 \pi M^{2}$ which are related to the properties of the unstable photon orbit \cite{Decanini:2011xw,Decanini:2011xi}. At high frequency approximation, it is shown that the oscillatory term arises from a sum of Regge poles which are characteristic resonances of the spacetime related to the QNMs. Briefly, Sinc approximation describes numerically and physically the fluctuations of the high energy absorption cross section.

The authors in Ref. \cite{Cavalcanti:2022cga} studied the behavior of one hairy black hole of the Ref. \cite{ovalle7} under scalar field perturbation, which proved QNM frequencies regulated by the hair. The main goal of this paper is to probe physical properties of another hairy black hole in Ref. \cite{ovalle7} caused by gravitational decoupling using quasinormal modes under massless scalar field, electromagnetic field and axial gravitational perturbations. This involves thorough calculations of how to see the effect of hair on QNM frequencies and time-domain profiles. The second goal is to study the greybody bound of hairy black hole caused by gravitational decoupling.
To that end, we show numerically how the hair appears in the greybody bounds, as well as high-energy absorption cross section with the Sinc approximation.

This paper is organized as follows. In the next section we brief review of hairy black hole caused by gravitational decoupling. Furthermore, we analyze the scalar field, electromagnetic field, and axial gravitational perturbations in the hairy black hole background spacetime. In Section \ref{QNMt}, the time-domain profiles of massless scalar field, electromagnetic field and axial gravitational perturbations in hairy black hole spacetime are given, and we also calculate the QNM frequencies using the Prony method and the 6th and 13th order WKB methods. In Section \ref{sec:level7}, the greybody factors and high-energy absorption cross section via sinc approximation was discussed. Section \ref{sec:summary} is a conclusion of the full text and a brief discussion of future directions.
\section{External fields and axial gravitational perturbations in hairy black hole spacetime}\label{bridf_review}
\subsection{A brief review of the hairy black hole}
Ovalle et al. used gravitational decoupling to give a spherically symmetric black hole with scalar hairs. This method has two characteristics: extending simple solutions to more complex fields and decoupling some complex sources of gravity. They assumed that the system has a well-defined event horizon and the conserved energy-momentum tensor describing the additional source satisfies the strong energy condition or dominant energy condition outside the event horizon \cite{ovalle7}. The metric has the form ($G = c = 1$)
\begin{equation}
d s^{2}=e^{\nu(r)} d t^{2}-e^{\lambda(r)} d r^{2}-r^{2}(d \theta^{2}+\sin ^{2} \theta d \phi^{2}),
\end{equation}
where
\begin{equation}\label{eq2}
e^{\nu}=e^{-\lambda}=1-\frac{2 \mathcal{M}}{r}+\alpha e^{-r /(\mathcal{M}-\alpha \l_0 / 2)},
\end{equation}
with the $\mathcal{M}$ is an asymptotic mass $\mathcal{M}=M+\alpha l_0 / 2$. This solution is presented by using gravitational decoupling and strong energy condition. Moreover, using the dominant energy condition (DEC), the ``charged" hairy black hole can be read as \cite{ovalle7}
\begin{equation}\label{eq3}
e^{\nu}=e^{-\lambda}=1-\frac{2 M+\alpha \l_0}{r}+\frac{Q^{2}}{r^{2}}-\frac{\alpha M e^{-r / M}}{r}.
\end{equation}
The solution (\ref{eq2}) has been studied in detail by Cavalcanti et al \cite{Cavalcanti:2022cga}. Therefore, all the content we give in the following is for the black hole (\ref{eq3}), and we will not discuss the relevant content of the black hole (\ref{eq2}). 
It is worth noting that we cannot reduce the hairy black hole (\ref{eq3}) to (\ref{eq2}) by setting $Q=0$. The main reason is that black hole solution (\ref{eq2}) is the Schwarzschild's deformation produced by a gravitational source which satisfies the strong energy conditions, while black hole solution (\ref{eq3}) is produced by a generic source satisfying the dominant energy conditions \cite{ovalle7}.
For the ``charged'' hairy black hole (\ref{eq3}), the event horizon radius $r_\mathrm{h}$ is given by the solution of the following equation
\begin{equation}
r_{\mathrm{h}}=\alpha l_0+2 M-\frac{Q^{2}}{r_{\mathrm{h}}}+\alpha M e^{-r_{\mathrm{h}} / M}.
\end{equation}

In addition, Ref. \cite{ovalle7} points out that the hairs $Q,\alpha,$ and $l_{0}$ of hairy black hole (\ref{eq3}) satisfy the following relationship
\begin{equation}
Q^{2} \geq 4 \alpha(M / e)^{2} \quad \text { and } \quad l_0 \geq M / e^{2}.
\end{equation}
From equation (5) we can see that the case of $Q=0$ and $\alpha$ non-zero is not allowed. Therefore, we do not discuss this case in this work. It should be noted that some regions of the parameter space will make hairy black holes (3) become a naked singularity. For example, when $Q>M$ with $a=0$, hairy black holes (3) become a naked singularity. Therefore, there should be a minimum $\alpha$ that turns the naked singularity into a black hole with an event horizon.

This hairy black hole (\ref{eq3}) contains parameters $Q, M, l_0, \alpha$, where $\{Q, \Delta=\alpha l_0\}$ denotes a potential
set of charges generating primary hair.
It should be noted that the charge $Q$ is not necessarily an electric charge of the Maxwell field source, because the source for the charge $Q$ is a tensor vacuum. The charge $Q$ could be a tidal charge from extra-dimensional origin or any other source. However, when charge $Q$ denotes an electric charge, one can say that the electro vacuum of the Reissner-Nordstr\"{o}m geometry also includes a tensor vacuum \cite{ovalle7}.
There are many similar examples. For example, the hair $Q$ is not an electric charge in Dadhich-Maartens-Papadopoulos-Rezania black hole solution \cite{Dadhich:2000am}, whose charge $Q$ is associated with a conformal theory.
In particular, the parameter $\Delta$ is closely related to the gauge transformation of the Schwarzschild metric, and it can push the event horizon to a place larger than the Schwarzschild radius, such that it can measure how much the black hole's entropy increases relative to the minimum Schwarzschild value as the black hole hair added \cite{ovalle7}.
\subsection{Scalar perturbation of the hair black hole}
In the context of general relativity, there are two  methods to study the black hole perturbations. The first method is to include a test field in the black hole background and to study the system by solving the dynamical equation for a specific test field in the black hole background. The second is the perturbation metric itself, i.e. the gravitational perturbation, in which the evolution equation is usually found by linearizing the Einstein equations. The most important feature between the two methods is that the gravitational radiation excited by gravitational perturbation is much stronger than the gravitational radiation excited by the perturbation of the external field of the black hole. We will first study the case where the external field is a scalar field. In curved spacetime, the perturbation equation of the massless scalar field can be written as
\begin{equation}
\frac{1}{\sqrt{-g}} \partial_{\mu}\left(\sqrt{-g} g^{\mu \nu} \partial_{\nu} \Psi\right)=0.
\end{equation}

Putting our considered hairy black hole metric into the scalar field equation, we can get

\begin{widetext}
\begin{equation}\label{eq9}
\begin{aligned}
-& \frac{\partial_{t}^{2} \Psi}{e^{\nu}}+\frac{1}{r^{2}}\left(2 r e^{\nu} \partial_{r} \Psi+r^{2} e^{\nu^{\prime}} \partial_{r} \Psi+r^{2} e^{\nu} \partial_{r}^{2} \Psi\right)+\frac{1}{r^{2}}\left(\frac{1}{\sin \theta} \partial_{\theta} \sin \theta \partial_{\theta} \Psi+\frac{1}{\sin ^{2} \theta} \partial_{\varphi}^{2} \Psi\right)=0,
\end{aligned}
\end{equation}
\end{widetext}
where $e^{\nu^{\prime}}$ denote $\frac{d}{d r} e^{\nu}$.
Due to the symmetry of the hair black hole spacetime background, we perform separation of variables
\begin{equation}
\Psi(t, r, \theta, \phi)=\sum_{l, m} \psi(t, r) Y_{l m}(\theta, \phi) / r,
\end{equation}
by substituting it into equation (\ref{eq9}), we can get the second order partial differential equation about the tortoise coordinate in the scalar field perturbation
\begin{equation}\label{eq11}
\frac{d \psi^{2}}{d t^{2}}-\frac{d \psi^{2}}{d \tau^{2}}+V(r) \psi=0,
\end{equation}
where tortoise coordinate $\tau$ is defined as $d\tau=1/e^{\nu} dr$, and $V(r)$ is effective potential
\begin{widetext}
\begin{equation}
V(r)=\left(1-\frac{2 M+\alpha \l_0}{r}+\frac{Q^{2}}{r^{2}}+\frac{\alpha M e^{-r / M}}{r}\right)\left[\frac{l(l+1)}{r^2}
 -\frac{2 Q^{2}}{r^{4}}+\frac{2 M}{r^{3}}+\frac{M \alpha e^{-r / M}}{r^{3}}+\frac{\alpha e^{-r / M}}{r^{2}}+\frac{\alpha l_{0}}{r^{3}}
\right].
\end{equation}
\end{widetext}
The perturbation of the scalar field has been receiving more attention. Moderski et al. have investigated the QNM spectrum under the perturbation of the scalar field in different black hole spacetimes \cite{Moderski:2001tk,Moderski:2005hf,Moderski:2001gt}. Boudet et al. have examined the QNM for Schwarzschild
black hole under projective invariant Chern-Simons modified gravity \cite{Boudet:2022wmb}. The authors in Ref. \cite{Franzin:2022iai} investigate QNM of scalar field perturbation in the spacetime of rotating regular black hole.

\subsection{Electromagnetic perturbation of the hair black hole}

The Maxwell equation satisfied by the electromagnetic field is
\begin{equation}\label{elec_mastereq}
\frac{1}{\sqrt{-g}} \frac{\partial}{\partial x^{\nu}}\left(\sqrt{-g} F^{\mu \nu}\right)=J^{\mu}.
\end{equation}
The electromagnetic field we are studying is a vacuum, so the value of the four-current density $J^{\mu}$ is 0.
$F^{\mu \nu}$ is the contravariant tensor of the electromagnetic field in the black hole spacetime background, and its covariant tensor is
\begin{equation}
F_{\mu \nu}=A_{\nu, \mu}-A_{\mu, \nu},
\end{equation}
where the comma denotes the ordinary derivative.
Since the background we are studying is spherically symmetric, $A_\mu$ can be expanded in the four-dimensional vector spherical harmonics as \cite{Cardoso:2001bb}
\begin{widetext}
\begin{equation}\label{eq13}
A_{\mu}(t, r, \theta, \phi)=\sum_{l, m}\left[\left(\begin{array}{c}
0 \\
0 \\
b^{l m}(t, r) \frac{1}{\sin \theta} \partial_{\phi} Y_{l m} \\
-b^{l m}(t, r) \sin \theta \partial_{\theta} Y_{l m}
\end{array}\right)+\left(\begin{array}{c}
f^{l m}(t, r) Y_{l m} \\
h^{l m}(t, r) Y_{l m} \\
k^{l m}(t, r) \partial_{\theta} Y_{l m} \\
k^{l m}(t, r) \partial_{\phi} Y_{l m}
\end{array}\right)\right],
\end{equation}
\end{widetext}
with $l$ being the angular quantum number, $m$ being the azimuthal number. The item on the left has parity $(-1)^{l+1}$, and the item on the right has parity $(-1)^{l}$. Moreover, $Y_{lm}$ is spherical harmonics.
Substituting equation (\ref{eq13}) into Maxwell's equation (\ref{elec_mastereq}), and using the tortoise coordinate, the second-order differential equation for the perturbation can be obtained
\begin{equation}
\frac{d \psi^{2}}{d t^{2}}-\frac{d \psi^{2}}{d \tau^{2}}+V(r) \psi=0,
\end{equation}
where the wavefunction $\psi(r)$ is a linear combination of $b^{l m}$, $f^{l m}$, $h^{l m}$ and $k^{l m}$, and the effective potential of the electromagnetic field perturbation in the hair black hole spacetime is
\begin{widetext}
\begin{equation}
V(r)=\left(1-\frac{2 M+\alpha \l_0}{r}+\frac{Q^{2}}{r^{2}}+\frac{\alpha M e^{-r / M}}{r}\right)\left[\frac{l(l+1)}{r^2}\right].
\end{equation}
\end{widetext}
%----------------------------------------------------------------%
\subsection{Axial gravitational perturbations of the hair black hole}

Regge and Wheeler were the first to study the axial gravitational perturbation of Schwarzschild black hole \cite{Regge:1957td}. The standard procedure for axial gravitational perturbation is to introduce a small perturbation $h_{\mu \nu}$ into the static background metric $g_{\mu \nu}^{0}$, i.e.
\begin{equation}
g_{\mu \nu}=g_{\mu \nu}^{0}+h_{\mu \nu},~ \left|h_{\mu \nu}\right| \ll 1.
\end{equation}
Moreover, the Ricci tensor can be expressed as
\begin{equation}
R_{\mu \nu}=R_{\mu \nu}^{0}+\delta R_{\mu \nu},
\end{equation}
with
\begin{equation}
\delta R_{\mu \nu}=\delta \Gamma_{\mu \alpha ; \nu}^{\alpha}-\delta \Gamma_{\mu \nu ; \alpha}^{\alpha},
\end{equation}
where
\begin{equation}
\delta \Gamma_{\mu \nu}^{\beta}=\frac{1}{2} g^{\beta \alpha}\left(h_{\alpha \nu ; \mu}+h_{\alpha \mu ; \nu}-h_{\mu \nu ; \alpha}\right).
\end{equation}
Regge and Wheeler proved that if the metric perturbation tensor $h_{\mu\nu}$ is expanded into tensor spherical harmonics, the equations describing the axial perturbations can be separated. Using the gauge symmetry of the field equations, one can obtain the general form of $h_{\mu\nu}$. The remaining components after simplification are \cite{Regge:1957td}
\begin{equation}
h_{\mu \nu}=\left(\begin{array}{cccc}
0 & 0 & 0 & h_{0}(t, r) \\
0 & 0 & 0 & h_{1}(t, r) \\
0 & 0 & 0 & 0 \\
h_{0}(t, r) & h_{1}(t, r) & 0 & 0
\end{array}\right) \sin \theta \partial_{\theta} P_{\ell}(\cos \theta) ,
\end{equation}
with $P_{\ell}(\cos \theta)$ being the Legendre polynomial of order $\ell$. Unknown functions $h_{0}(t, r)$ and $h_{1}(t, r)$ satisfy the following equations
\begin{widetext}
\begin{equation}
\frac{1}{f(r)} \frac{d h_{0}}{d t}-\frac{df(r) }{d r}h_{1}
-\frac{d h_{1}}{d r}f(r)=0,
\end{equation}
\begin{equation}
\left(\frac{d^{2} h_{1}}{d t^{2}}-\frac{d^{2} h_{0}}{d t d r}+\frac{2}{r} \frac{d h_{0}}{d t}\right)\frac{1}{f(r)}+\frac{l^2+l-2}{r^{2}}h_{1}=0,
\end{equation}
\begin{equation}
\left(\frac{d^{2} h_{0}}{d r^{2}}-\frac{d^{2} h_{1}}{d t d r}-\frac{2}{r} \frac{d h_{1}}{d t}\right)\frac{f(r)}{2}+\frac{h_{0}}{r^{2}}\left[r \frac{d f(r)}{d r}-\frac{1}{2} l(l+1)\right] =0,
\end{equation}
\end{widetext}
where $f(r)$ is the $g_{tt}$ in the spherical symmetry metric.
By defining $\Psi(t, r)=(1 / r) f(r) h_{1}(t, r)$, and according to the tortoise coordinate, we can gain the Schr\"{o}dinger-like equation of the axial gravitational perturbation
\begin{equation}
\frac{d \Psi^{2}}{d t^{2}}-\frac{d \Psi^{2}}{d \tau^{2}}+V(r) \Psi=0,
\end{equation}
where the Regge-Wheeler potential $V(r)$ can be read as
\begin{widetext}
\begin{equation}
V(r)=\left(1-\frac{2 M+\alpha \l_0}{r}+\frac{Q^{2}}{r^{2}}+\frac{\alpha M e^{-r / M}}{r}\right)\left[\frac{l(l+1)}{r^2}+\frac{6 Q^{2}}{r^{4}}-\frac{6 M}{r^{3}}-\frac{3 M \alpha e^{-r/M}}{r^{3}}-\frac{3 \alpha e^{-r/M}}{r^{2}}-\frac{3 \alpha l_{0}}{r^{3}}\right].
\end{equation}
\end{widetext}

\section{QNM of hair black hole}\label{QNMt}
In this section, we will numerically solve the wave equation in hairy black hole spacetime background to gain the time-domain profiles of this spacetime. In addition, we use the Prony method to extract quasinormal modes, and our results are in good agreement with those calculated by higher-order WKB. To obtain the time-domain profiles of the hairy black hole, we introduce the light cone coordinates
\begin{equation}
\begin{array}{l}
u=t-\tau, \\
v=t+\tau,
\end{array}
\end{equation}
then Eq. (\ref{eq11}) can be written as
\begin{equation}\label{eq15}
\frac{\partial^{2}}{\partial u \partial v} \psi(u, v)=-\frac{1}{4}V(r) \psi(u, v).
\end{equation}

The Eq. (\ref{eq15}) can be discreted as \cite{price,price2}
\begin{equation}
\psi_{N}=\psi_{E}+\psi_{W}-\psi_{S}-\delta u \delta v V\left(\frac{\psi_{W}+\psi_{E}}{8}\right)+O\left(\Delta^{4}\right).
\end{equation}
where $S=(u,v),W=(u+\delta u,v),E=(u,v+\delta v),N=(u+\delta u,v+\delta v)$.
Moreover, we use the following initial Gaussian pulse
\begin{equation}
\begin{array}{l}
\psi\left(u=u_{0}, v\right)=\exp \left[-\frac{\left(v-v_{c}\right)^{2}}{2 \sigma^{2}}\right], \\
\psi\left(u, v=v_{0}\right)=0,
\end{array}
\end{equation}
where $u_{0}=v_{0}=0,\sigma=3,v_{c}=10$.
As thus, the time-domain profiles of hairy black hole caused by gravitational decoupling can be obtained.

\begin{figure}[t!]\centering
%\flushleft
%\centering 
%\setlength{\unitlength}{1.5cm}
%\hspace{0.4cm}
\vspace{0.6cm}
\includegraphics[scale=0.6]{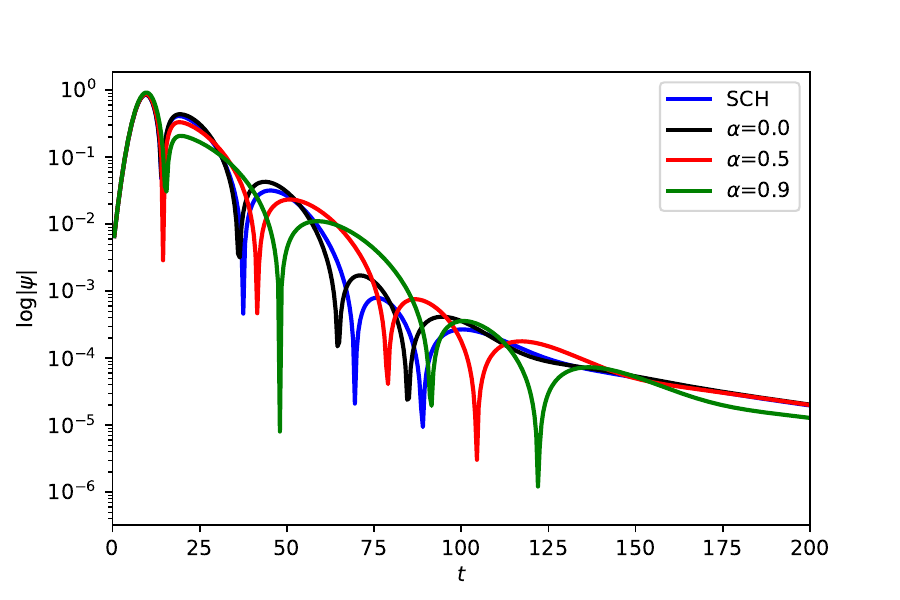}
\includegraphics[scale=0.6]{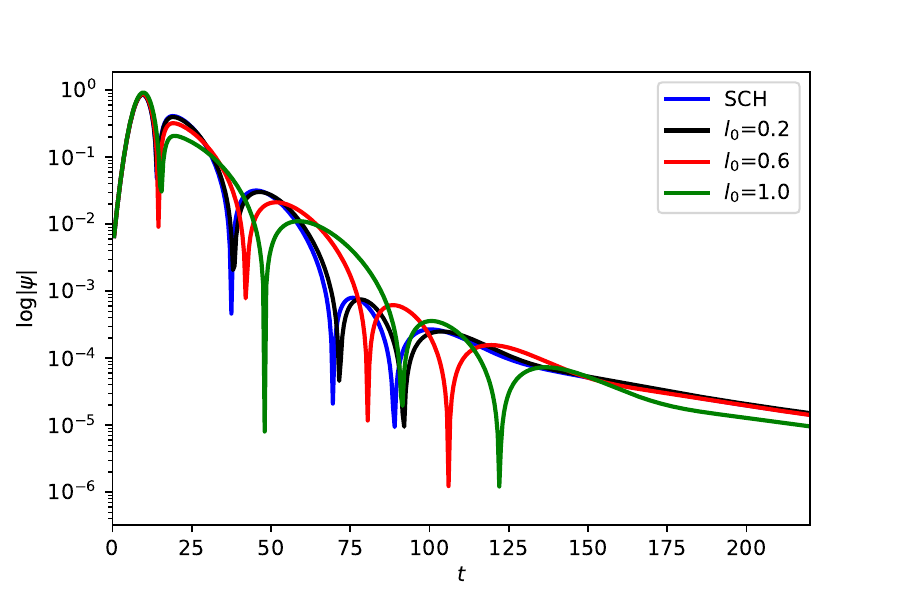}
\includegraphics[scale=0.6]{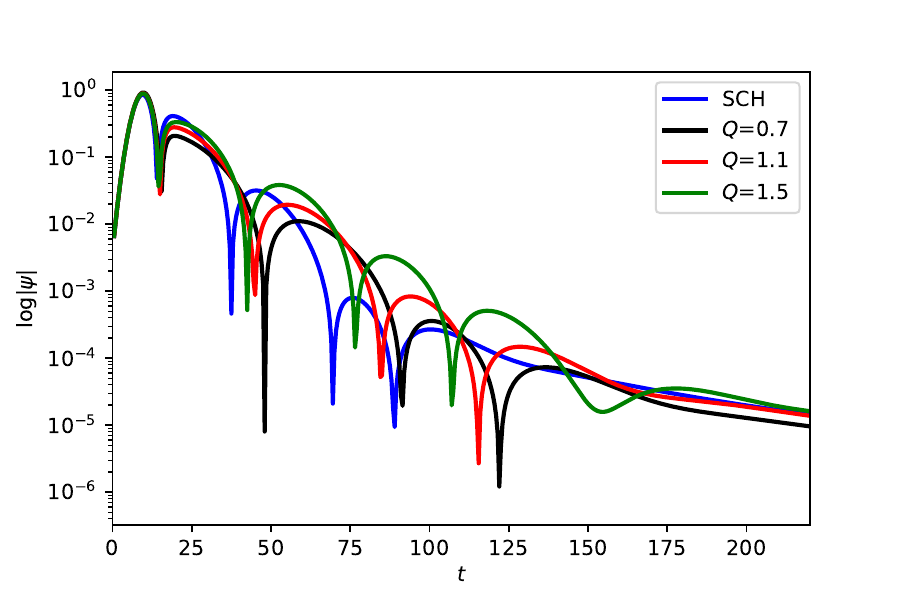}
\setlength{\abovecaptionskip}{-0.1cm}
\setlength{\belowcaptionskip}{0.8cm}
\caption{The time-domain profiles of the scalar field perturbation in the hairy black hole for different $\alpha$ with $M=1,l=0,l_0=1,Q=0.7$, for different $l_0$ with $M=1,l=0,\alpha=0.9,Q=0.7$, and for different $Q$ with $M=1,l=0,l_0=1,\alpha=0.9$, respectively. The blue line represents the TDF of the Schwarzschild black hole (SCH) under scalar field perturbation with $M=1,l=0$.}
\label{scalar_l0}
\vspace{0.1cm}
\end{figure}

\begin{figure}[t!]\centering
\begin{center}
\vspace{0.6cm}
\includegraphics[scale=0.6]{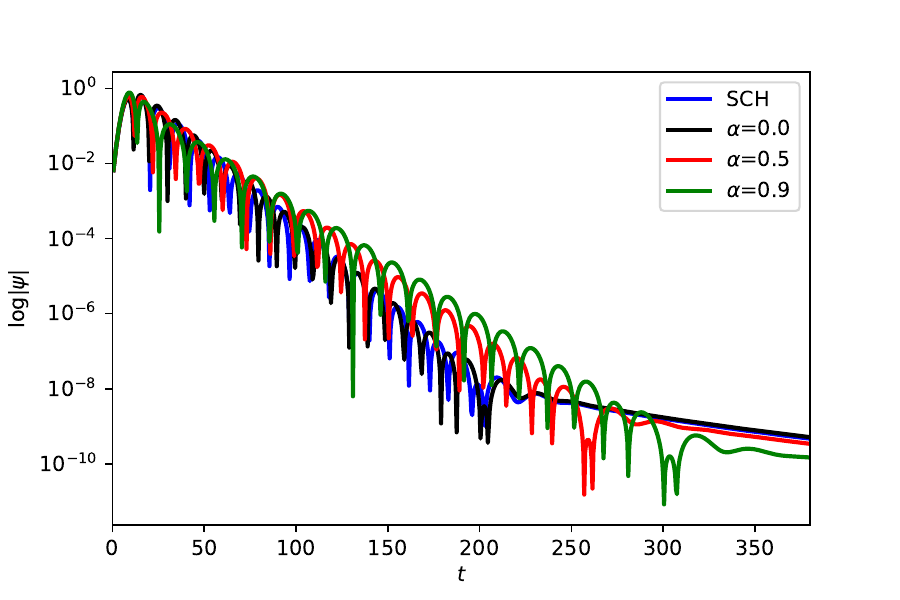}
\includegraphics[scale=0.6]{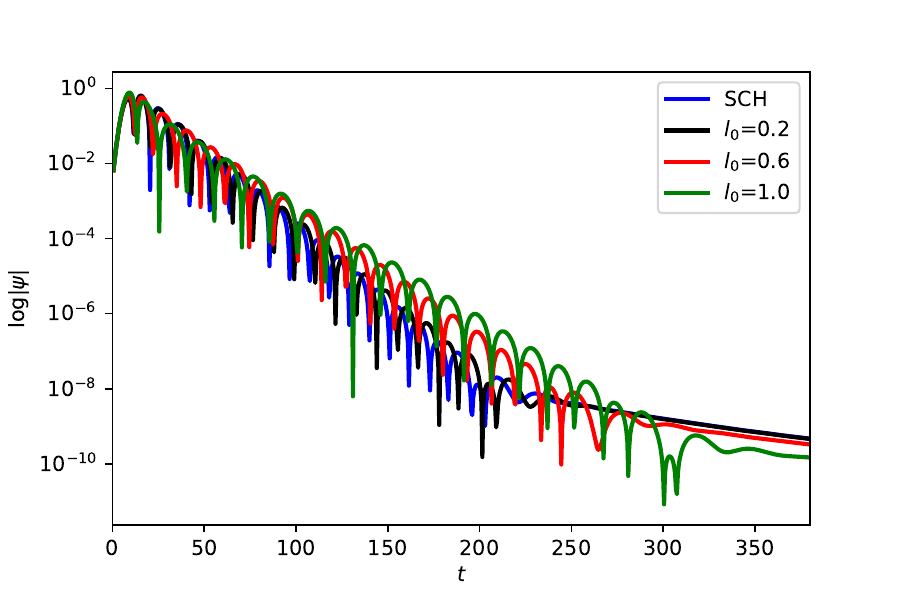}
\includegraphics[scale=0.6]{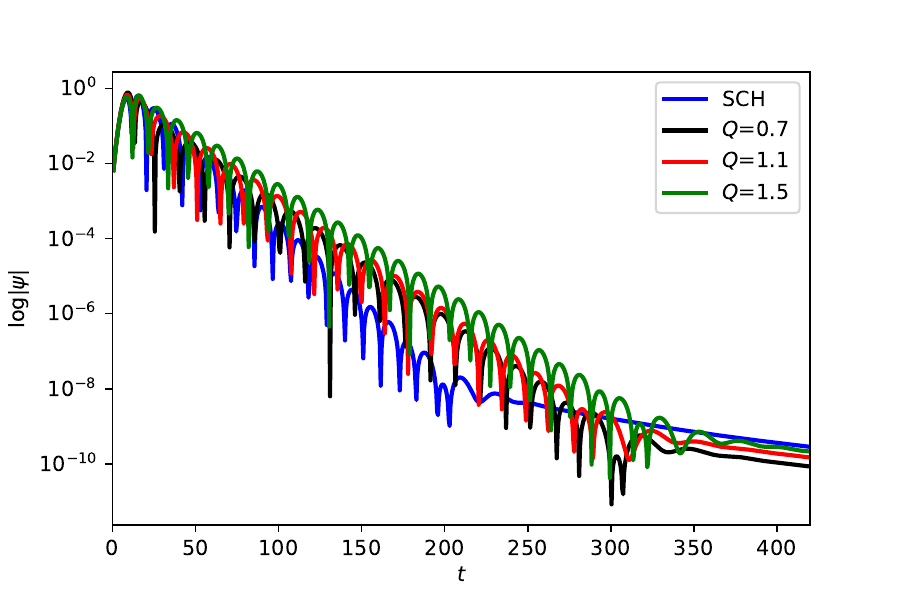}
\end{center}
\setlength{\abovecaptionskip}{-0.1cm}
\setlength{\belowcaptionskip}{0.8cm}
\caption{The time-domain profiles of the scalar field perturbation in the hairy black hole for different $\alpha$ with $M=1,l=1,l_0=1,Q=0.7$, for different $l_0$ with $M=1,l=1,\alpha=0.9,Q=0.7$, and for different $Q$ with $M=1,l=1,l_0=1,\alpha=0.9$, respectively. The blue line represents the TDF of the Schwarzschild black hole (SCH) under scalar field perturbation with $M=1,l=1$.}
\label{scalar}
\end{figure}

In Fig. \ref{scalar_l0} and \ref{scalar}, we show the time-domain profiles (TDF) of the scalar field perturbation in the hairy black hole. The difference between these two figures is that we study the case of $l=0$ in the former, and the case of $l=1$ in the latter. In the top panel of Fig. \ref{scalar_l0} and Fig. \ref{scalar}, we show the influence of the black hole parameter $\alpha$ on TDF, where $M=1,l_0=1,Q=0.7$. In addition, we also give the TDF of the Schwarzschild black hole (SCH). One can find that when $\alpha$ is the smallest, its damped is the strongest. In other words, the decay rate is the fastest when $\alpha$ is the smallest, which implies that the imaginary part of the QNM frequency will decrease as $\alpha$ increases.
In the middle panel of Fig. \ref{scalar_l0} and Fig. \ref{scalar}, we give the influence of hairy black hole charge $l_0$ on its TDF with $M=1,\alpha=0.9,Q=0.7$.
It shows that as $l_0$ increases, TDF damping slower, which is an indication that the imaginary part of the QNM frequency that denotes damping should also be decreasing when charge $l_0$ increases.
In the bottom panel of Fig. \ref{scalar_l0} and Fig. \ref{scalar}, the influence of the charge $Q$ on the hairy black hole is presented, where $M=1, \alpha=0.9, l_0=1$. Our results show that when the $Q$ is larger, its TDF decays more slowly. Moreover, in these two figures one can find that the clear power-law tail.

\begin{figure}[ht!]\centering
\hspace{0.5cm}
\begin{center}
\includegraphics[scale=0.6]{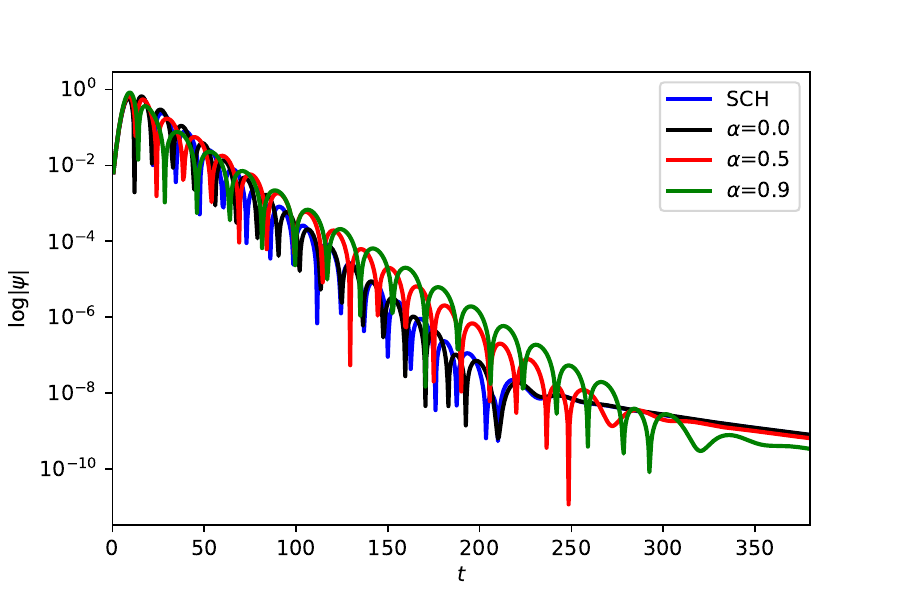}
\includegraphics[scale=0.6]{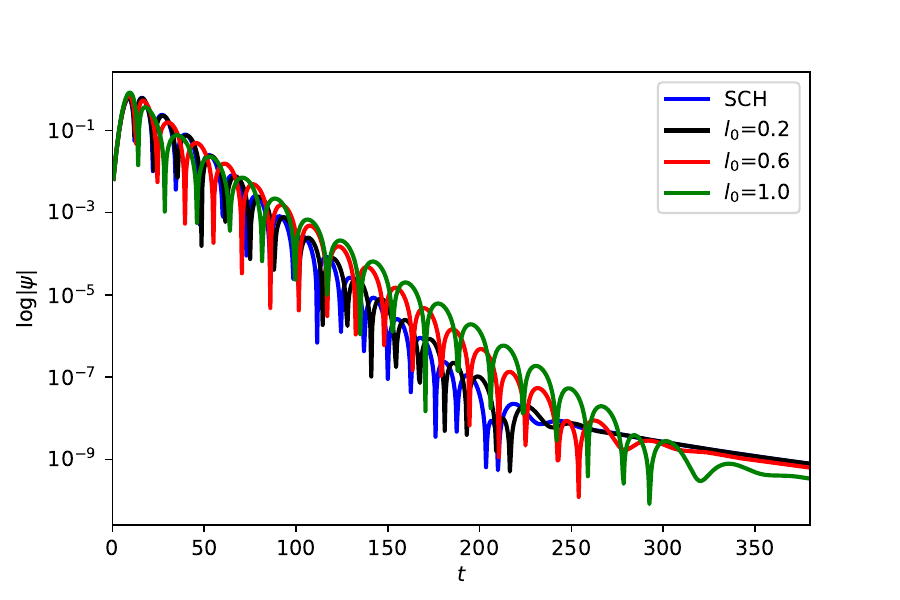}
\includegraphics[scale=0.6]{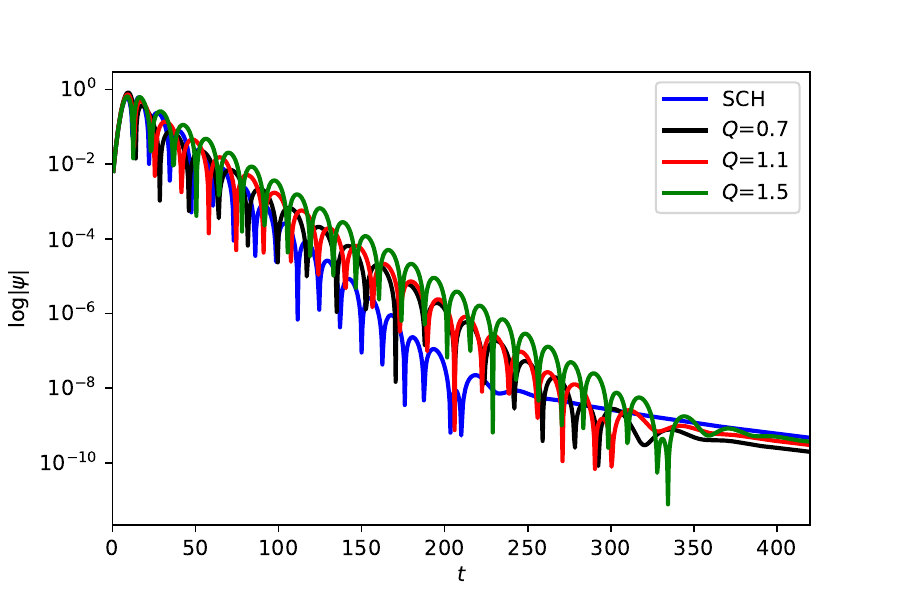}
\end{center}
\setlength{\abovecaptionskip}{-0.1cm}
\setlength{\belowcaptionskip}{0.8cm}
\caption{The time-domain profiles of the electromagnetic field perturbation in the hairy black hole for different $\alpha$ with $M=1,l=1,l_0=1,Q=0.7$, for different $l_0$ with $M=1,l=1,\alpha=0.9,Q=0.7$, and for different $Q$ with $M=1,l=1,l_0=1,\alpha=0.9$, respectively. The blue line represents the TDF of the Schwarzschild black hole (SCH) under electromagnetic field perturbation with $M=1, l=1$.}
\label{elec}
\end{figure}

In Fig. \ref{elec}, the TDF of the electromagnetic field in the hair black hole are given. Form Fig. \ref{elec}, we find that larger $\alpha$, $l_0$ and $Q$ make TDF decay slower. The contribution of these parameters are similar to those of the scalar field perturbation, which demonstrates that electromagnetic field perturbation and scalar field perturbation make the TDF of hairy black hole exhibit similar behavior.

\begin{figure}[ht!]\centering
%\setlength{\unitlength}{1.5cm}
%\begin{center}
%\hspace{-0.3cm}
\vspace{0.5cm}
\includegraphics[scale=0.6]{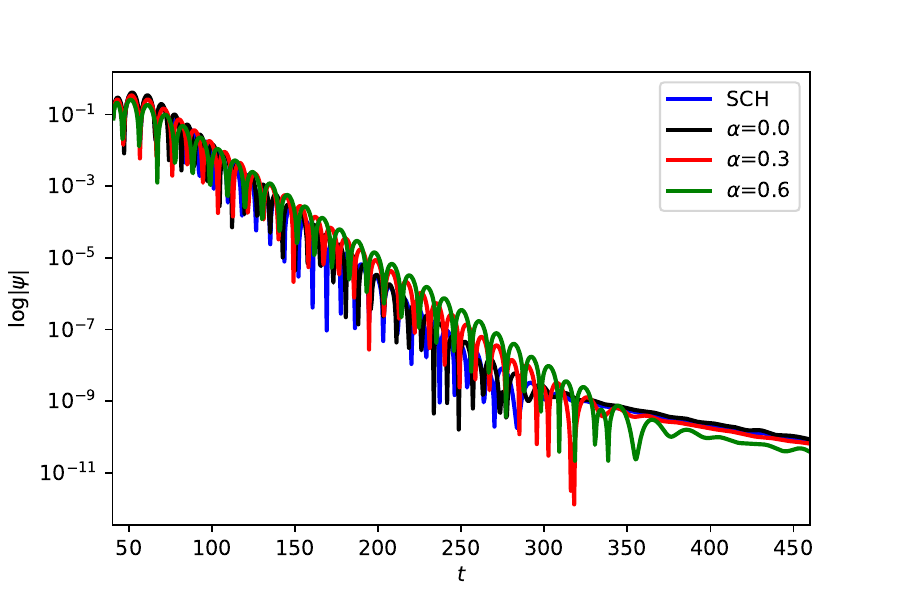}
\includegraphics[scale=0.6]{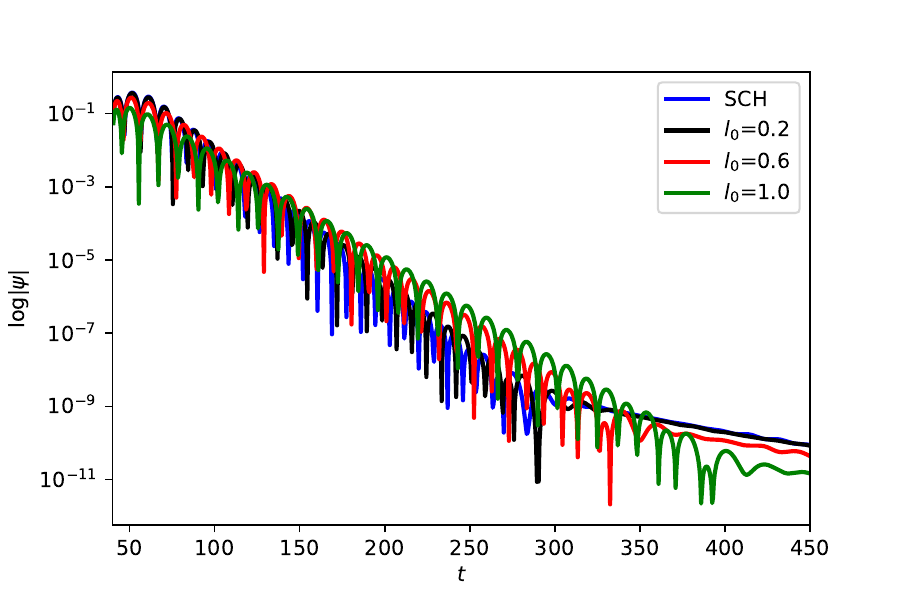}
\includegraphics[scale=0.6]{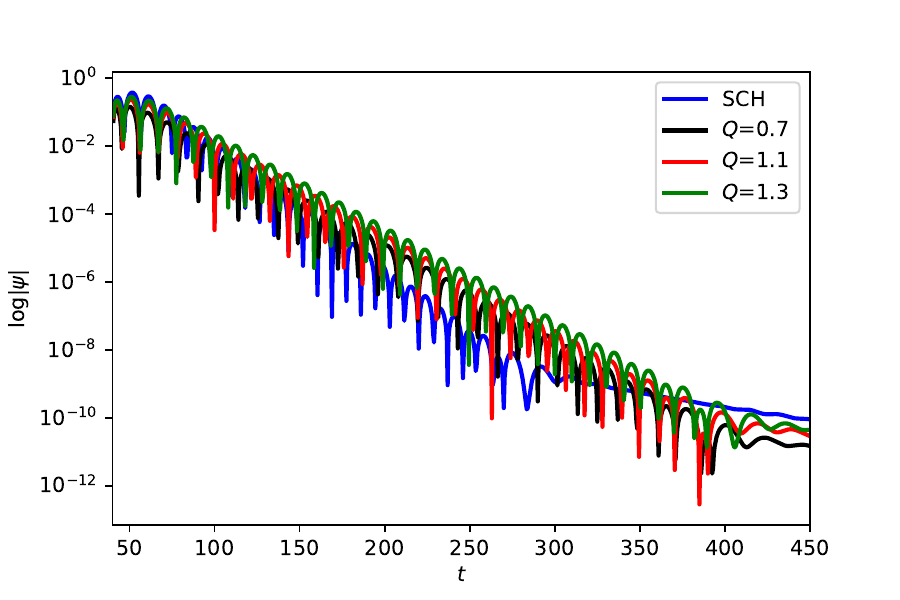}
%\end{center}
\setlength{\abovecaptionskip}{-0.1cm}
\setlength{\belowcaptionskip}{0.8cm}
\caption{The time-domain profiles of the gravitational perturbation  in the hairy black hole for different $\alpha$ with $M=1,l=2,l_0=1,Q=0.7$, for different $l_0$ with $M=1,l=2,\alpha=0.9,Q=0.7$, and for different $Q$ with $M=1,l=2,l_0=1,\alpha=0.9$, respectively. The blue line represents the TDF of the Schwarzschild black hole (SCH) under scalar field perturbation with $M=1, l=2$.}
\label{grav}
\vspace{0.3cm}
\end{figure}

In Fig. \ref{grav}, we present the TDF of the gravitational perturbation in the hairy black hole. The influence of $\alpha$, $l_0$ and charge $Q$ on hairy black hole are studied respectively. We can see that the effects of $\alpha$, $l_0$ and $Q$ on the hairy black hole we study under the gravitational perturbation are similar to the scalar field perturbation and the electromagnetic field perturbation.
%But the most obvious difference between the TDF under gravitational perturbation and the TDF under scalar field perturbation and electromagnetic field perturbation is that there is no power-law tail, which may be due to the fact that gravitational radiation excited by gravitational perturbations is much stronger than the gravitational radiation excited by the perturbation of the external field of the hairy black hole.This may also be one of the characteristics of hairy black holes under strong gravitational radiation.

Apart from studying the time-domain profile, we also studied the QNM frequency. Through continuous development, many methods have been proposed to calculate QNM frequency. The P\"{o}shl-Teller potential approximation method \cite{pt1,pt2} is an earlier method for calculating the black hole's QNM frequency, but only when the effective potential of the black hole is similar to the P\"{o}shl-Teller potential can a more accurate result be obtained. In Ref. \cite{Cavalcanti:2022cga}, the authors used this method to calculate the QNM frequencies of hairy black hole under scalar perturbation. In Ref. \cite{Konoplya:2003ii,Konoplya:2010kv}, Konoplya uses the 6th order WKB method to study the scattering problem. In Ref. \cite{Matyjasek:2017psv}, 13th order WKB method was presented by using Pad\'{e} approximation. In our work, we use the WKB method \cite{wkb1,wkb2,wkb4,hwkb1} and the Prony method to study the QNM frequencies of hairy black hole caused by gravitational decoupling.
The Prony method extracts the QNM frequency from the time-domain profiles by the damped exponents \cite{prony,Konoplya:2011qq}
\begin{equation}
\phi(t) \simeq \sum_{i=1}^{p} C_{i} e^{-i \omega_{i} t}.
\end{equation}
We will compare the QNM frequencies extracted from the time-domain profiles using the Prony method with the results calculated by the higher-order WKB method to verify the correctness of time-domain profiles.
We have listed the value of QNM frequencies in TABLE \ref{scaltablew}, TABLE \ref{electablew}, TABLE \ref{gravtablew} and TABLE \ref{diffl}.

The QNM frequencies of scalar field perturbation for hairy black hole are presented in TABLE \ref{scaltablew}.
We have studied three cases in TABLE \ref{scaltablew}, the first one is to fix other parameters ($M=1,l=1,l_0=1,Q=0.7$) to study the effect of $\alpha$ on the QNM frequencies of hairy black hole, the second one is to specify other parameters ($M=1,l=1,\alpha=0.9,Q=0.7$) to research the impact of $l_0$ on the QNM frequencies of hairy black hole, and the third one is to fixed other parameters ($M=1,l=1,l_0=1,\alpha=0.9$) to investigate the effect of $Q$ on the QNM frequencies of hairy black hole.
One can see that there is only a slight deviation between QNM frequency extracted by TDF and the result of high-order WKB in three cases, which is enough to prove the accuracy of TDF. It is seen that when $\alpha$ increases, the real part and imaginary part of QNM frequency are both smaller, that is, the oscillation frequency of GW is decreasing and the damping is also decreasing. In addition, when $l_0$ increases, we can also find that the real and imaginary parts of QNM frequency are decreasing. The imaginary part decrease indicates that its decay slows down, which is consistent with the results shown in Fig. \ref{scalar}.
However, we find that the same trend for different $\alpha$ and $l_0$ does not continue to the results for various charges $Q$.
The increase of charge $Q$ under the scalar perturbation leads to the real part of QNM frequencies increasing, whereas the imaginary part decreases, which demonstrates that the oscillation frequency of GW is increasing, and the decay rate is slower. This is the reason why the power-law tails appear early in Fig. \ref{scalar} when charge $Q$ is small, and this behavior is similar to the results of different $\alpha$ and $l_0$, i.e. smaller $\alpha$ and $l_0$ correspond to earlier power-law tails.

In TABLE \ref{electablew}, the QNM frequencies of electromagnetic field perturbation for hairy black hole are given. For different $\alpha$, the parameter settings are
$M=1,l=1,l_0=1,Q=0.7$.
For different $l_0$, the parameter settings are
$M=1,l=1,\alpha=0.9,Q=0.7$.
For different $Q$, the parameter settings are
$M=1,l=1,l_0=1,\alpha=0.9$.
In addition, we also give the QNM frequencies of hairy black hole under gravitational perturbation in TABLE \ref{gravtablew}.
We can see that TABLE \ref{electablew} and TABLE \ref{gravtablew} present the same trend as TABLE \ref{scaltablew} for different cases, which further demonstrates that scalar field perturbation, electromagnetic perturbation and gravitational perturbation have similar qualitative behavior to the hairy black hole caused by gravitational decoupling. In particular, the effects of $\alpha$ and $l_0$ on the real part of QNM are opposite to one of $Q$, i.e., the increase of $\alpha$ and $l_0$ will decrease the oscillation frequency of gravitational wave, whereas the increase of $Q$ will increase the oscillation frequency of gravitational wave.

The results of TABLE \ref{diffl} are QNM frequencies of scalar field,  electromagnetic field, and gravitational perturbations with different $l$. It is interesting to note that the increase of the multipole moment $l$ will significantly increase the oscillation frequency of gravitational waves, whereas the effect on the decay rate is very small. The most important point in TABLE \ref{diffl} is that when $\alpha=0, Q=0$, our results can reproduce the results of the Schwarzschild black hole very well. Moreover, in TABLE \ref{diffl}, we calculate the error of 13th order WKB method. One can find that the error of 13th order WKB method for a small $l$ will be larger than that of the large $l$, and the error of scalar field perturbation is larger than that of electromagnetic field and gravitational perturbation. For example, for the gravitational perturbation results in TABLE \ref{diffl}, when $l=2$, the error is about $4.47532\times10^{-5}$, and about $1.03191\times10^{-8}$ for $l=4$. On the other hand, TABLE \ref{table5} gives the deviation of the QNM frequencies of Schwarzschild black hole ($M=1,\alpha=0,Q=0$) calculated by the Prony method relative to Iyer's results \cite{Iyer:1986nq}. We can see that for the scalar field perturbation $l=0$, the deviation of decay rate reaches 6.9\%, and for other cases, whether it is for frequency or decay rate, the deviation does not exceed 2.7\%.
The reason for the large deviation of scalar field perturbation $l = 0$ may be found in Fig. \ref{scalar_l0}. We can see that the duration of the ringdown of TDF is very short from Fig. \ref{scalar_l0}, which implies that the Prony method is not particularly accurate in this case.
In TABLE \ref{table6}, the behaviour from the real parts and imaginary parts of QNM frequencies for hairy black hole under gravitational perturbation with different overtone numbers are showed. We observe that the real parts of QNM frequencies decrease with the overtone number, and the imaginary parts of QNM frequencies increase with the overtone number.

\begin{table*}[htbp!]
\renewcommand\arraystretch{1.3}
\caption{QNM frequencies of scalar field perturbation for hairy black hole.}
\setlength{\abovecaptionskip}{0.5cm}
\setlength{\belowcaptionskip}{0.5cm}
\setlength{\tabcolsep}{3mm}{
\begin{tabular}{|cccc|}
\hline
\multicolumn{4}{|c|}{$M=1,l=1,l_0=1,Q=0.7$}                                                                                                                  \\ \hline
\multicolumn{1}{|c|}{$\alpha$} & \multicolumn{1}{c|}{Prony method $(M\omega)$} & \multicolumn{1}{c|}{6th order WKB $(M\omega)$} & 13th order WKB $(M\omega)$ \\ \hline
\multicolumn{1}{|c|}{0.0}      & \multicolumn{1}{c|}{0.321507 - 0.101701$i$}     & \multicolumn{1}{c|}{0.322775 - 0.0994117$i$}      & 0.322772 - 0.0993483$i$       \\ \hline
\multicolumn{1}{|c|}{0.1}      & \multicolumn{1}{c|}{0.300244 - 0.0964229$i$}     & \multicolumn{1}{c|}{0.303041 - 0.0951931$i$}      & 0.303039 - 0.0951213$i$       \\ \hline
\multicolumn{1}{|c|}{0.2}      & \multicolumn{1}{c|}{0.282813 - 0.091302$i$}     & \multicolumn{1}{c|}{0.286089 - 0.091189$i$}       & 0.286092 - 0.0911003$i$       \\ \hline
\multicolumn{1}{|c|}{0.3}      & \multicolumn{1}{c|}{0.268623 - 0.0878727$i$}      & \multicolumn{1}{c|}{0.271271 - 0.0874315$i$}      & 0.27128 - 0.0873223$i$        \\ \hline
\multicolumn{1}{|c|}{0.4}      & \multicolumn{1}{c|}{0.255067 - 0.0820306$i$}       & \multicolumn{1}{c|}{0.258144 - 0.0839217$i$}      & 0.258165 - 0.0837874$i$       \\ \hline
\multicolumn{1}{|c|}{0.5}      & \multicolumn{1}{c|}{0.243646 - 0.0790852$i$}     & \multicolumn{1}{c|}{0.246392 - 0.0806478$i$}      & 0.246423 - 0.0804828$i$       \\ \hline
\multicolumn{1}{|c|}{0.6}      & \multicolumn{1}{c|}{0.233439 - 0.0769056$i$}     & \multicolumn{1}{c|}{0.23578 - 0.0775934$i$}       & 0.235838 - 0.0773884$i$       \\ \hline
\multicolumn{1}{|c|}{0.7}      & \multicolumn{1}{c|}{0.224392 - 0.0749798$i$}     & \multicolumn{1}{c|}{0.226129 - 0.0747407$i$}      & 0.226207 - 0.0744979$i$       \\ \hline
\multicolumn{1}{|c|}{0.8}      & \multicolumn{1}{c|}{0.217691 - 0.0723938$i$}     & \multicolumn{1}{c|}{0.217300 - 0.072072$i$}       & 0.217378 - 0.0718025$i$       \\ \hline
\multicolumn{1}{|c|}{0.9}      & \multicolumn{1}{c|}{0.209181 - 0.0695709$i$}     & \multicolumn{1}{c|}{0.209181 - 0.0695709$i$}      & 0.209304 - 0.0692642$i$       \\ \hline
\multicolumn{4}{|c|}{$M=1,l=1,\alpha=0.9,Q=0.7$}                                                                                                             \\ \hline
\multicolumn{1}{|c|}{$l_0$}    & \multicolumn{1}{c|}{Prony method $(M\omega)$} & \multicolumn{1}{c|}{6th order WKB $(M\omega)$} & 13th order WKB $(M\omega)$ \\ \hline
\multicolumn{1}{|c|}{0.2}      & \multicolumn{1}{c|}{0.281022 - 0.0955926$i$}     & \multicolumn{1}{c|}{0.283488 - 0.0947429$i$}      & 0.283471 - 0.0945504$i$       \\ \hline
\multicolumn{1}{|c|}{0.3}      & \multicolumn{1}{c|}{0.269242 - 0.0913397$i$}     & \multicolumn{1}{c|}{0.27147 - 0.0906598$i$}       & 0.271478 - 0.0904487$i$       \\ \hline
\multicolumn{1}{|c|}{0.4}      & \multicolumn{1}{c|}{0.258385 - 0.0874759$i$}     & \multicolumn{1}{c|}{0.260416 - 0.0869074$i$}      & 0.260451 - 0.0866649$i$       \\ \hline
\multicolumn{1}{|c|}{0.5}      & \multicolumn{1}{c|}{0.24837 - 0.0839461$i$}      & \multicolumn{1}{c|}{0.250217 - 0.0834488$i$}      & 0.250271 - 0.083192$i$        \\ \hline
\multicolumn{1}{|c|}{0.6}      & \multicolumn{1}{c|}{0.239092 - 0.0807142$i$}     & \multicolumn{1}{c|}{0.240779 - 0.0802517$i$}      & 0.240844 - 0.0799829$i$       \\ \hline
\multicolumn{1}{|c|}{0.7}      & \multicolumn{1}{c|}{0.23046 - 0.0777224$i$}     & \multicolumn{1}{c|}{0.232021 - 0.0772882$i$}      & 0.232096 - 0.0769971$i$       \\ \hline
\multicolumn{1}{|c|}{0.8}      & \multicolumn{1}{c|}{0.222408 - 0.0749569$i$}      & \multicolumn{1}{c|}{0.223874 - 0.0745339$i$}      & 0.223966 - 0.074229$i$        \\ \hline
\multicolumn{1}{|c|}{0.9}      & \multicolumn{1}{c|}{0.214889 - 0.0723729$i$}     & \multicolumn{1}{c|}{0.216278 - 0.0719677$i$}      & 0.216403 - 0.071666$i$        \\ \hline
\multicolumn{1}{|c|}{1.0}      & \multicolumn{1}{c|}{0.209181 - 0.0695709$i$}     & \multicolumn{1}{c|}{0.209181 - 0.0695709$i$}      & 0.209304 - 0.0692642$i$       \\ \hline
\multicolumn{4}{|c|}{$M=1,l=1,l_0=1,\alpha=0.9$}                                                                                                             \\ \hline
\multicolumn{1}{|c|}{$Q$}      & \multicolumn{1}{c|}{Prony method $(M\omega)$} & \multicolumn{1}{c|}{6th order WKB $(M\omega)$} & 13th order WKB $(M\omega)$ \\ \hline
\multicolumn{1}{|c|}{0.7}      & \multicolumn{1}{c|}{0.209181 - 0.0695709$i$}     & \multicolumn{1}{c|}{0.209181 - 0.0695709$i$}      & 0.209304 - 0.0692642$i$       \\ \hline
\multicolumn{1}{|c|}{0.8}      & \multicolumn{1}{c|}{0.21196 - 0.0707743$i$}     & \multicolumn{1}{c|}{0.212052 - 0.0698582$i$}      & 0.212157 - 0.069565$i$        \\ \hline
\multicolumn{1}{|c|}{0.9}      & \multicolumn{1}{c|}{0.214681 - 0.0711001$i$}     & \multicolumn{1}{c|}{0.21552 - 0.0701646$i$}       & 0.215628 - 0.0699185$i$       \\ \hline
\multicolumn{1}{|c|}{1.0}      & \multicolumn{1}{c|}{0.218256 - 0.0709727$i$}     & \multicolumn{1}{c|}{0.219709 - 0.0704705$i$}      & 0.219804 - 0.0702317$i$       \\ \hline
\multicolumn{1}{|c|}{1.1}      & \multicolumn{1}{c|}{0.222866 - 0.0706441$i$}      & \multicolumn{1}{c|}{0.224788 - 0.0707406$i$}      & 0.224854 - 0.0705321$i$       \\ \hline
\multicolumn{1}{|c|}{1.2}      & \multicolumn{1}{c|}{0.228786 - 0.0700971$i$}     & \multicolumn{1}{c|}{0.231015 - 0.0709032$i$}      & 0.231059 - 0.070716$i$        \\ \hline
\multicolumn{1}{|c|}{1.3}      & \multicolumn{1}{c|}{0.23683 - 0.0712016$i$}     & \multicolumn{1}{c|}{0.238783 - 0.0708064$i$}      & 0.238807 - 0.070646$i$        \\ \hline
\multicolumn{1}{|c|}{1.4}      & \multicolumn{1}{c|}{0.246787 - 0.0708725$i$}     & \multicolumn{1}{c|}{0.248804 - 0.070053$i$}       & 0.248757 - 0.0699446$i$       \\ \hline
\multicolumn{1}{|c|}{1.5}      & \multicolumn{1}{c|}{0.262831 - 0.0696549$i$}     & \multicolumn{1}{c|}{0.262188 - 0.0674775$i$}      & 0.261999 - 0.0674607$i$       \\ \hline
\end{tabular}}
\label{scaltablew}
\end{table*}

\begin{table*}[]
\renewcommand\arraystretch{1.3}
\setlength{\abovecaptionskip}{0.2cm}
\setlength{\belowcaptionskip}{0.3cm}
\caption{QNM frequencies of electromagnetic field perturbation for hairy black hole.}
\setlength{\tabcolsep}{3mm}{
\begin{tabular}{|cccc|}
\hline
\multicolumn{4}{|c|}{$M=1,l=1,l_0=1,Q=0.7$}                                                                                                                  \\ \hline
\multicolumn{1}{|c|}{$\alpha$} & \multicolumn{1}{c|}{Prony method $(M\omega)$} & \multicolumn{1}{c|}{6th order WKB $(M\omega)$} & 13th order WKB $(M\omega)$ \\ \hline
\multicolumn{1}{|c|}{0.0}      & \multicolumn{1}{c|}{0.27438 - 0.0952652$i$}  & \multicolumn{1}{c|}{0.277264 - 0.0951071$i$}   & 0.277328 - 0.0948974$i$    \\ \hline
\multicolumn{1}{|c|}{0.1}      & \multicolumn{1}{c|}{0.257552 - 0.0915071$i$}   & \multicolumn{1}{c|}{0.259522 - 0.0909144$i$}   & 0.259592 - 0.0907009$i$    \\ \hline
\multicolumn{1}{|c|}{0.2}      & \multicolumn{1}{c|}{0.242204 - 0.0868901$i$}  & \multicolumn{1}{c|}{0.244445 - 0.0869722$i$}   & 0.244524 - 0.0867263$i$    \\ \hline
\multicolumn{1}{|c|}{0.3}      & \multicolumn{1}{c|}{0.229607 - 0.0834999$i$}  & \multicolumn{1}{c|}{0.231379 - 0.0832965$i$}   & 0.231462 - 0.0830699$i$    \\ \hline
\multicolumn{1}{|c|}{0.4}      & \multicolumn{1}{c|}{0.218479 - 0.0802702$i$}  & \multicolumn{1}{c|}{0.21989 - 0.0798778$i$}    & 0.219977 - 0.0796442$i$    \\ \hline
\multicolumn{1}{|c|}{0.5}      & \multicolumn{1}{c|}{0.208539 - 0.0771759$i$}  & \multicolumn{1}{c|}{0.209666 - 0.0766987$i$}   & 0.20978 - 0.0764508$i$     \\ \hline
\multicolumn{1}{|c|}{0.6}      & \multicolumn{1}{c|}{0.199608 - 0.0742378$i$}  & \multicolumn{1}{c|}{0.200482 - 0.0737389$i$}   & 0.200604 - 0.0734884$i$    \\ \hline
\multicolumn{1}{|c|}{0.7}      & \multicolumn{1}{c|}{0.191541 - 0.0714914$i$}  & \multicolumn{1}{c|}{0.192165 - 0.0709789$i$}   & 0.192297 - 0.0707073$i$    \\ \hline
\multicolumn{1}{|c|}{0.8}      & \multicolumn{1}{c|}{0.184193 - 0.0689691$i$} & \multicolumn{1}{c|}{0.184584 - 0.0684005$i$}   & 0.184723 - 0.0681492$i$    \\ \hline
\multicolumn{1}{|c|}{0.9}      & \multicolumn{1}{c|}{0.177388 - 0.0672283$i$}  & \multicolumn{1}{c|}{0.177635 - 0.0659867$i$}   & 0.17778 - 0.0657295$i$     \\ \hline
\multicolumn{4}{|c|}{$M=1,l=1,\alpha=0.9,Q=0.7$}                                                                                                             \\ \hline
\multicolumn{1}{|c|}{$l_0$}    & \multicolumn{1}{c|}{Prony method $(M\omega)$} & \multicolumn{1}{c|}{6th order WKB $(M\omega)$} & 13th order WKB $(M\omega)$ \\ \hline
\multicolumn{1}{|c|}{0.2}      & \multicolumn{1}{c|}{0.238957 - 0.0903121$i$}   & \multicolumn{1}{c|}{0.240451 - 0.0899668$i$}   & 0.240592 - 0.0895844$i$    \\ \hline
\multicolumn{1}{|c|}{0.3}      & \multicolumn{1}{c|}{0.228902 - 0.086411$i$}   & \multicolumn{1}{c|}{0.230301 - 0.086083$i$}    & 0.230435 - 0.0857165$i$    \\ \hline
\multicolumn{1}{|c|}{0.4}      & \multicolumn{1}{c|}{0.219661 - 0.0827886$i$}   & \multicolumn{1}{c|}{0.220963 - 0.0825113$i$}   & 0.221115 - 0.082122$i$     \\ \hline
\multicolumn{1}{|c|}{0.5}      & \multicolumn{1}{c|}{0.211118 - 0.0794722$i$}  & \multicolumn{1}{c|}{0.212344 - 0.0792167$i$}   & 0.212507 - 0.0788714$i$    \\ \hline
\multicolumn{1}{|c|}{0.6}      & \multicolumn{1}{c|}{0.203217 - 0.0763974$i$}  & \multicolumn{1}{c|}{0.204367 - 0.0761695$i$}   & 0.204528 - 0.075829$i$     \\ \hline
\multicolumn{1}{|c|}{0.7}      & \multicolumn{1}{c|}{0.195489 - 0.0730517$i$}  & \multicolumn{1}{c|}{0.196962 - 0.0733442$i$}   & 0.197118 - 0.0730409$i$    \\ \hline
\multicolumn{1}{|c|}{0.8}      & \multicolumn{1}{c|}{0.189112 - 0.0709165$i$}  & \multicolumn{1}{c|}{0.190071 - 0.070718$i$}    & 0.190225 - 0.0704139$i$    \\ \hline
\multicolumn{1}{|c|}{0.9}      & \multicolumn{1}{c|}{0.182467 - 0.0680648$i$}  & \multicolumn{1}{c|}{0.183643 - 0.0682712$i$}   & 0.183795 - 0.0679941$i$    \\ \hline
\multicolumn{1}{|c|}{1.0}      & \multicolumn{1}{c|}{0.177388 - 0.0672283$i$}   & \multicolumn{1}{c|}{0.177635 - 0.0659867$i$}   & 0.17778 - 0.0657295$i$     \\ \hline
\multicolumn{4}{|c|}{$M=1,l=1,l_0=1,\alpha=0.9$}                                                                                                             \\ \hline
\multicolumn{1}{|c|}{$Q$}      & \multicolumn{1}{c|}{Prony method $(M\omega)$} & \multicolumn{1}{c|}{6th order WKB $(M\omega)$} & 13th order WKB $(M\omega)$ \\ \hline
\multicolumn{1}{|c|}{0.7}      & \multicolumn{1}{c|}{0.177388 - 0.0672283$i$}   & \multicolumn{1}{c|}{0.177635 - 0.0659867$i$}   & 0.17778 - 0.0657295$i$     \\ \hline
\multicolumn{1}{|c|}{0.8}      & \multicolumn{1}{c|}{0.179783 - 0.0668075$i$}  & \multicolumn{1}{c|}{0.180372 - 0.0663587$i$}   & 0.180517 - 0.066101$i$     \\ \hline
\multicolumn{1}{|c|}{0.9}      & \multicolumn{1}{c|}{0.18274 - 0.0668985$i$}  & \multicolumn{1}{c|}{0.183697 - 0.0667678$i$}   & 0.183843 - 0.066497$i$     \\ \hline
\multicolumn{1}{|c|}{1.0}      & \multicolumn{1}{c|}{0.186791 - 0.0674179$i$}  & \multicolumn{1}{c|}{0.18774 - 0.0671974$i$}    & 0.187859 - 0.066929$i$     \\ \hline
\multicolumn{1}{|c|}{1.1}      & \multicolumn{1}{c|}{0.191665 - 0.067882$i$}     & \multicolumn{1}{c|}{0.192683 - 0.0676172$i$}      & 0.192789 - 0.0673285$i$       \\ \hline
\multicolumn{1}{|c|}{1.2}      & \multicolumn{1}{c|}{0.197621 - 0.0682103$i$}     & \multicolumn{1}{c|}{0.198814 - 0.0679582$i$}      & 0.198927 - 0.0676624$i$       \\ \hline
\multicolumn{1}{|c|}{1.3}      & \multicolumn{1}{c|}{0.205651 - 0.0687884$i$}      & \multicolumn{1}{c|}{0.206573 - 0.0680707$i$}      & 0.206693 - 0.0677617$i$       \\ \hline
\multicolumn{1}{|c|}{1.4}      & \multicolumn{1}{c|}{0.216046 - 0.0686647$i$}     & \multicolumn{1}{c|}{0.216777 - 0.0675363$i$}      & 0.216889 - 0.0672306$i$       \\ \hline
\multicolumn{1}{|c|}{1.5}      & \multicolumn{1}{c|}{0.229584 - 0.0662932$i$}  & \multicolumn{1}{c|}{0.230929 - 0.0649291$i$}   & 0.230986 - 0.0646869$i$    \\ \hline
\end{tabular}}
\label{electablew}
\vspace{0.7cm}
\end{table*}

\begin{table*}[htbp!]
\renewcommand\arraystretch{1.3}
\setlength{\abovecaptionskip}{0.2cm}
\setlength{\belowcaptionskip}{0.3cm}
\caption{QNM frequencies of gravitational perturbation for hairy black hole.}
\setlength{\tabcolsep}{3mm}{
\begin{tabular}{|cccc|}
\hline
\multicolumn{4}{|c|}{$M=1,l=2,l_0=1,Q=0.7$}                                                                                                                  \\ \hline
\multicolumn{1}{|c|}{$\alpha$} & \multicolumn{1}{c|}{Prony method $(M\omega)$} & \multicolumn{1}{c|}{6th order WKB $(M\omega)$} & 13th order WKB $(M\omega)$ \\ \hline
\multicolumn{1}{|c|}{0.0}      & \multicolumn{1}{c|}{0.421319 - 0.093321 $i$}  & \multicolumn{1}{c|}{0.41962 - 0.0910065$i$}   & 0.419637 - 0.0909727$i$    \\ \hline
\multicolumn{1}{|c|}{0.1}      & \multicolumn{1}{c|}{0.394301 - 0.0872158$i$}  & \multicolumn{1}{c|}{0.392428 - 0.087046$i$}   & 0.392426 - 0.0870356$i$    \\ \hline
\multicolumn{1}{|c|}{0.2}      & \multicolumn{1}{c|}{0.369882 - 0.085768$i$}  & \multicolumn{1}{c|}{0.369425 - 0.0833178$i$}   & 0.369409 - 0.0833276$i$     \\ \hline
\multicolumn{1}{|c|}{0.3}      & \multicolumn{1}{c|}{0.350611 - 0.0815108$i$}   & \multicolumn{1}{c|}{0.349553 - 0.0798375$i$}   & 0.349535 - 0.079867$i$    \\ \hline
\multicolumn{1}{|c|}{0.4}      & \multicolumn{1}{c|}{0.332492 - 0.0763707$i$}  & \multicolumn{1}{c|}{0.332112 - 0.0765966$i$}   & 0.332093 - 0.0766124$i$    \\ \hline
\multicolumn{1}{|c|}{0.5}      & \multicolumn{1}{c|}{0.315724 - 0.0736334$i$}  & \multicolumn{1}{c|}{0.316614 - 0.0735789$i$}   & 0.316588 - 0.0736032$i$    \\ \hline
\multicolumn{1}{|c|}{0.6}      & \multicolumn{1}{c|}{0.303406 - 0.0710666$i$}  & \multicolumn{1}{c|}{0.302703 - 0.0707663$i$}   & 0.30268 - 0.0707763$i$    \\ \hline
\multicolumn{1}{|c|}{0.7}      & \multicolumn{1}{c|}{0.290561 - 0.0682911$i$}   & \multicolumn{1}{c|}{0.290115 - 0.0681411$i$}   & 0.290095 - 0.0681618$i$    \\ \hline
\multicolumn{1}{|c|}{0.8}      & \multicolumn{1}{c|}{0.278566 - 0.066302$i$}  & \multicolumn{1}{c|}{0.278644 - 0.0656872$i$}   & 0.27864 - 0.0657188$i$    \\ \hline
\multicolumn{1}{|c|}{0.9}      & \multicolumn{1}{c|}{0.266795 - 0.0643959$i$}  & \multicolumn{1}{c|}{0.26813 - 0.0633896$i$}   & 0.268131 - 0.0634126$i$    \\ \hline
\multicolumn{4}{|c|}{$M=1,l=2,\alpha=0.9,Q=0.7$}                                                                                                             \\ \hline
\multicolumn{1}{|c|}{$l_0$}    & \multicolumn{1}{c|}{Prony method $(M\omega)$} & \multicolumn{1}{c|}{6th order WKB $(M\omega)$} & 13th order WKB $(M\omega)$ \\ \hline
\multicolumn{1}{|c|}{0.2}      & \multicolumn{1}{c|}{0.364607 - 0.0865609$i$}  & \multicolumn{1}{c|}{0.363216 - 0.0864224$i$}   & 0.363004 - 0.0863902$i$    \\ \hline
\multicolumn{1}{|c|}{0.3}      & \multicolumn{1}{c|}{0.346775 - 0.0833835$i$}  & \multicolumn{1}{c|}{0.347865 - 0.08270$i$}   & 0.347729 - 0.0827333$i$    \\ \hline
\multicolumn{1}{|c|}{0.4}      & \multicolumn{1}{c|}{0.333299 - 0.0791465$i$}   & \multicolumn{1}{c|}{0.333738 - 0.0792706$i$}     & 0.333645 - 0.0793122$i$    \\ \hline
\multicolumn{1}{|c|}{0.5}      & \multicolumn{1}{c|}{0.320798 - 0.0759211$i$}  & \multicolumn{1}{c|}{0.320696 - 0.0761044$i$}   & 0.320636 - 0.0761492$i$      \\ \hline
\multicolumn{1}{|c|}{0.6}      & \multicolumn{1}{c|}{0.309047 - 0.0731385$i$}  & \multicolumn{1}{c|}{0.30862 - 0.0731748$i$}   & 0.308578 - 0.0732138$i$     \\ \hline
\multicolumn{1}{|c|}{0.7}      & \multicolumn{1}{c|}{0.298027 - 0.0706296$i$}  & \multicolumn{1}{c|}{0.297408 - 0.0704583$i$}   & 0.29738 - 0.0704918$i$     \\ \hline
\multicolumn{1}{|c|}{0.8}      & \multicolumn{1}{c|}{0.287702 - 0.0683308$i$}   & \multicolumn{1}{c|}{0.286972 - 0.0679339$i$}   & 0.286956 - 0.0679646$i$    \\ \hline
\multicolumn{1}{|c|}{0.9}      & \multicolumn{1}{c|}{0.278041 - 0.0662711$i$}   & \multicolumn{1}{c|}{0.277235 - 0.0655832$i$}   & 0.277228 - 0.0656096$i$    \\ \hline
\multicolumn{1}{|c|}{1.0}      & \multicolumn{1}{c|}{0.266795 - 0.0643959$i$}  & \multicolumn{1}{c|}{0.26813 - 0.0633896$i$}   & 0.268131 - 0.0634126$i$    \\ \hline
\multicolumn{4}{|c|}{$M=1,l=2,l_0=1,\alpha=0.9$}                                                                                                             \\ \hline
\multicolumn{1}{|c|}{$Q$}      & \multicolumn{1}{c|}{Prony method $(M\omega)$} & \multicolumn{1}{c|}{6th order WKB $(M\omega)$} & 13th order WKB $(M\omega)$ \\ \hline
\multicolumn{1}{|c|}{0.7}      & \multicolumn{1}{c|}{0.266795 - 0.0643959$i$}  & \multicolumn{1}{c|}{0.26813 - 0.0633896$i$}   & 0.268131 - 0.0634126$i$    \\ \hline
\multicolumn{1}{|c|}{0.8}      & \multicolumn{1}{c|}{0.271865 - 0.0645973$i$}  & \multicolumn{1}{c|}{0.272466 - 0.0637162$i$}    & 0.272459 - 0.0637328$i$    \\ \hline
\multicolumn{1}{|c|}{0.9}      & \multicolumn{1}{c|}{0.278479 - 0.0651984$i$}  & \multicolumn{1}{c|}{0.277738 - 0.0640789$i$}   & 0.27774 - 0.0640953$i$    \\ \hline
\multicolumn{1}{|c|}{1.0}      & \multicolumn{1}{c|}{0.284881 - 0.0651774$i$}  & \multicolumn{1}{c|}{0.284153 - 0.0644651$i$}   & 0.284135 - 0.0644731$i$     \\ \hline
\multicolumn{1}{|c|}{1.1}      & \multicolumn{1}{c|}{0.292663 - 0.0653426$i$}     & \multicolumn{1}{c|}{0.292016 - 0.0648472$i$}      & 0.292003 - 0.0648387$i$       \\ \hline
\multicolumn{1}{|c|}{1.2}      & \multicolumn{1}{c|}{0.302629 - 0.0659483$i$}     & \multicolumn{1}{c|}{0.301795 - 0.065162$i$}      & 0.301803 - 0.0651248$i$       \\ \hline
\multicolumn{1}{|c|}{1.3}      & \multicolumn{1}{c|}{0.315112 - 0.0655522$i$}     & \multicolumn{1}{c|}{0.314282 - 0.0652534$i$}      & 0.314329 - 0.0651797$i$       \\ \hline
\multicolumn{1}{|c|}{1.4}      & \multicolumn{1}{c|}{0.33039 - 0.0639798$i$}     & \multicolumn{1}{c|}{0.330944 - 0.0646919$i$}      & 0.331053 - 0.0645997$i$       \\ \hline
\multicolumn{1}{|c|}{1.5}      & \multicolumn{1}{c|}{0.354654 - 0.0600052$i$}  & \multicolumn{1}{c|}{0.355086 - 0.0619111$i$}   & 0.355174 - 0.0618972$i$    \\ \hline
\end{tabular}}
\label{gravtablew}
\vspace{0.7cm}
\end{table*}

\begin{table*}[htbp!]
\setlength{\abovecaptionskip}{0.2cm}
\setlength{\belowcaptionskip}{0.3cm}
\caption{QNM frequencies of scalar field (Scal),  electromagnetic field (Elec), and gravitational (Grav) perturbations for different $l$.}
\renewcommand\arraystretch{1.3}
\setlength{\tabcolsep}{2.5mm}{
\begin{tabular}{|cccccc|}
\hline
\multicolumn{6}{|c|}{$M=1,\alpha=0,Q=0$}                                                                                                                                                                                           \\ \hline
\multicolumn{1}{|c|}{Field}                 & \multicolumn{1}{l|}{$l$} & \multicolumn{1}{c|}{Prony method $(M\omega)$}  & \multicolumn{1}{c|}{6th order WKB $(M\omega)$} & \multicolumn{1}{c|}{13th order WKB $(M\omega)$} & Error \\ \hline
\multicolumn{1}{|c|}{\multirow{3}{*}{Scal}} & \multicolumn{1}{l|}{0}   & \multicolumn{1}{c|}{0.105932 - 0.103975$i$}     & \multicolumn{1}{c|}{0.110493 - 0.100793$i$}    & \multicolumn{1}{c|}{0.111336 - 0.103793$i$}     &   $2.57859\times 10^{-3}$    \\ \cline{2-6}
\multicolumn{1}{|c|}{}                      & \multicolumn{1}{l|}{1}   & \multicolumn{1}{c|}{0.289254 - 0.0967834$i$}   & \multicolumn{1}{c|}{0.29291 - 0.0977616$i$}    & \multicolumn{1}{c|}{0.292935 - 0.0976625$i$}    &   $5.63328\times 10^{-6}$    \\ \cline{2-6}
\multicolumn{1}{|c|}{}                      & \multicolumn{1}{l|}{2}   & \multicolumn{1}{c|}{0.476682 - 0.0964193$i$}   & \multicolumn{1}{c|}{0.483642 - 0.0967661$i$}   & \multicolumn{1}{c|}{0.483643 - 0.0967596$i$}    &   $3.07067\times 10^{-7}$    \\ \hline
\multicolumn{1}{|c|}{\multirow{3}{*}{Elec}} & \multicolumn{1}{l|}{1}   & \multicolumn{1}{c|}{0.245525 - 0.0912522$i$}   & \multicolumn{1}{c|}{0.248191 - 0.092637$i$}    & \multicolumn{1}{c|}{0.24826 - 0.0924874$i$}     &   $1.65754\times 10^{-5}$    \\ \cline{2-6}
\multicolumn{1}{|c|}{}                      & \multicolumn{1}{l|}{2}   & \multicolumn{1}{c|}{0.451157 - 0.0927749$i$}   & \multicolumn{1}{c|}{0.457593 - 0.095011$i$}    & \multicolumn{1}{c|}{0.457595 - 0.0950047$i$}    &   $4.82647\times 10^{-7}$    \\ \cline{2-6}
\multicolumn{1}{|c|}{}                      & \multicolumn{1}{l|}{3}   & \multicolumn{1}{c|}{0.644805 - 0.0930214$i$}   & \multicolumn{1}{c|}{0.656898 - 0.0956171$i$}   & \multicolumn{1}{c|}{0.656899 - 0.0956163$i$}    &   $4.67008\times 10^{-7}$    \\ \hline
\multicolumn{1}{|c|}{\multirow{3}{*}{Grav}} & \multicolumn{1}{l|}{2}   & \multicolumn{1}{c|}{0.369919 - 0.0893064$i$}     & \multicolumn{1}{c|}{0.373619 - 0.088891$i$}    & \multicolumn{1}{c|}{0.373583 - 0.0889827$i$}    &   $8.39293\times 10^{-5}$    \\ \cline{2-6}
\multicolumn{1}{|c|}{}                      & \multicolumn{1}{l|}{3}   & \multicolumn{1}{c|}{0.590055 - 0.0919906$i$}   & \multicolumn{1}{c|}{0.599443 - 0.0927025$i$}   & \multicolumn{1}{c|}{0.599443 - 0.0927028$i$}    &   $1.79319\times 10^{-8}$    \\ \cline{2-6}
\multicolumn{1}{|c|}{}                      & \multicolumn{1}{l|}{4}   & \multicolumn{1}{c|}{0.792386 - 0.094910$i$}   & \multicolumn{1}{c|}{0.809178 - 0.0941641$i$}   & \multicolumn{1}{c|}{0.809178 - 0.094164$i$}     &   $3.42659\times 10^{-8}$    \\ \hline
\multicolumn{6}{|c|}{$M=1,l_0=1,\alpha=0.9,Q=0.7$}                                                                                                                                                                                 \\ \hline
\multicolumn{1}{|c|}{Field}                 & \multicolumn{1}{l|}{$l$} & \multicolumn{1}{c|}{Prony method $(M\omega)$}  & \multicolumn{1}{c|}{6th order WKB $(M\omega)$} & \multicolumn{1}{c|}{13th order WKB $(M\omega)$} & Error \\ \hline
\multicolumn{1}{|c|}{\multirow{3}{*}{Scal}} & \multicolumn{1}{l|}{0}   & \multicolumn{1}{c|}{0.0757766 - 0.0719676$i$}  & \multicolumn{1}{c|}{0.0850839 - 0.064694$i$}   & \multicolumn{1}{c|}{0.0840857 - 0.0674921$i$}   &   $4.27326\times 10^{-3}$       \\ \cline{2-6}
\multicolumn{1}{|c|}{}                      & \multicolumn{1}{l|}{1}   & \multicolumn{1}{c|}{0.209635 - 0.0696289$i$}   & \multicolumn{1}{c|}{0.209181 - 0.0695709$i$}   & \multicolumn{1}{c|}{0.209304 - 0.0692642$i$}    &   $1.35598\times 10^{-5}$    \\ \cline{2-6}
\multicolumn{1}{|c|}{}                      & \multicolumn{1}{l|}{2}   & \multicolumn{1}{c|}{0.345137 - 0.0681247$i$}   & \multicolumn{1}{c|}{0.345759 - 0.0687022$i$}   & \multicolumn{1}{c|}{0.345765 - 0.0686858$i$}    &   $4.57878\times 10^{-7}$    \\ \hline
\multicolumn{1}{|c|}{\multirow{3}{*}{Elec}} & \multicolumn{1}{l|}{1}   & \multicolumn{1}{c|}{0.177388 - 0.0672283$i$}    & \multicolumn{1}{c|}{0.177635 - 0.0659867$i$}   & \multicolumn{1}{c|}{0.17778 - 0.0657295$i$}     &   $1.51868\times 10^{-5}$    \\ \cline{2-6}
\multicolumn{1}{|c|}{}                      & \multicolumn{1}{l|}{2}   & \multicolumn{1}{c|}{0.325863 - 0.0671239$i$}   & \multicolumn{1}{c|}{0.327385 - 0.0675054$i$}   & \multicolumn{1}{c|}{0.327391 - 0.0674944$i$}    &   $1.26352\times 10^{-6}$    \\ \cline{2-6}
\multicolumn{1}{|c|}{}                      & \multicolumn{1}{l|}{3}   & \multicolumn{1}{c|}{0.466492 - 0.0689451$i$} & \multicolumn{1}{c|}{0.469885 - 0.0679249$i$}   & \multicolumn{1}{c|}{0.469885 - 0.0679233$i$}    &     $1.61534\times 10^{-7}$  \\ \hline
\multicolumn{1}{|c|}{\multirow{3}{*}{Grav}} & \multicolumn{1}{l|}{2}   & \multicolumn{1}{c|}{0.266795 - 0.0643959$i$}   & \multicolumn{1}{c|}{0.26813 - 0.0633896$i$}   & \multicolumn{1}{c|}{0.268131 - 0.0634126$i$}    &    $4.47532\times 10^{-5}$   \\ \cline{2-6}
\multicolumn{1}{|c|}{}                      & \multicolumn{1}{l|}{3}   & \multicolumn{1}{c|}{0.425910 - 0.066640$i$}   & \multicolumn{1}{c|}{0.429374 - 0.0659638$i$}    & \multicolumn{1}{c|}{0.429374 - 0.0659635$i$}    &    $7.92364\times 10^{-8}$   \\ \cline{2-6}
\multicolumn{1}{|c|}{}                      & \multicolumn{1}{l|}{4}   & \multicolumn{1}{c|}{0.580939 - 0.067654$i$}   & \multicolumn{1}{c|}{0.579217 - 0.0669528$i$}   & \multicolumn{1}{c|}{0.579217 - 0.0669526$i$}    &   $1.03191\times 10^{-8}$    \\ \hline
\end{tabular}  }
\label{diffl}
\end{table*}
 %------------------------------------------------------------------%

\begin{table*}[htbp!]
\setlength{\abovecaptionskip}{0.2cm}
\setlength{\belowcaptionskip}{0.3cm}
\caption{Deviation of the QNM frequencies of Schwarzschild black hole ($M=1,\alpha=0,Q=0$) calculated by the Prony method relative to Iyer's results.}
\renewcommand\arraystretch{1.3}
\setlength{\tabcolsep}{2.5mm}{
\begin{tabular}{|c|c|c|c|c|c|}
\hline
Field                 & $l$ & Prony method $(M\omega)$ & Iyer' Results\cite{Iyer:1986nq} $(M\omega)$ & Frequency'deviations & Decay rate'deviations \\ \hline
\multirow{3}{*}{Scal} & 0   & 0.103129 - 0.107250$i$   & 0.1046 - 0.1152$i$      & 1.4\%                & 6.9\%                 \\ \cline{2-6}
                      & 1   & 0.289254 - 0.0967834$i$  & 0.2911 - 0.0980$i$      & 0.6\%                & 1.2\%                 \\ \cline{2-6}
                      & 2   & 0.476682 - 0.0964193$i$  & 0.4832 - 0.0968$i$      & 1.3\%                & 0.4\%                 \\ \hline
\multirow{3}{*}{Elec} & 1   & 0.245525 - 0.0912522$i$  & 0.2459 - 0.0931$i$      & 0.2\%                & 2.0\%                 \\ \cline{2-6}
                      & 2   & 0.451157 - 0.0927749$i$  & 0.4571 - 0.0951$i$      & 1.3\%                & 2.4\%                 \\ \cline{2-6}
                      & 3   & 0.644805 - 0.0930214$i$  & 0.6567 - 0.0956$i$      & 1.8\%                & 2.7\%                 \\ \hline
\multirow{3}{*}{Grav} & 2   & 0.369919 - 0.0893064$i$  & 0.3732 - 0.0892$i$      & 0.9\%                & 0.1\%                 \\ \cline{2-6}
                      & 3   & 0.590055 - 0.0919906$i$  & 0.5993 - 0.0927$i$      & 1.5\%                & 0.7\%                 \\ \cline{2-6}
                      & 4   & 0.792386 - 0.094910$i$    & 0.8091 - 0.0942$i$      & 2.0\%                & 0.8\%                 \\ \hline
\end{tabular}}
\label{table5}
\end{table*}

\begin{table*}[htbp!]
\setlength{\abovecaptionskip}{0.2cm}
\setlength{\belowcaptionskip}{0.3cm}
\caption{The QNM frequencies of hairy black hole for gravitational perturbation with different overtone numbers.}
\renewcommand\arraystretch{1.3}
\setlength{\tabcolsep}{2.5mm}{
\begin{tabular}{|ccccccccc|}
\hline
\multicolumn{9}{|c|}{$\alpha=0.01,Q=0.075,l_0=0.15$}                                                                                                                                                                                                                                                                                 \\ \hline
\multicolumn{1}{|c|}{$l$}                & \multicolumn{1}{c|}{$n$} & \multicolumn{1}{c|}{$(M\omega_n)$}          & \multicolumn{1}{c|}{$l$}                & \multicolumn{1}{c|}{$n$} & \multicolumn{1}{c|}{$(M\omega_n)$}           & \multicolumn{1}{c|}{$l$}                & \multicolumn{1}{c|}{$n$} & $(M\omega_n)$           \\ \hline
\multicolumn{1}{|c|}{\multirow{4}{*}{2}} & \multicolumn{1}{c|}{0}   & \multicolumn{1}{c|}{0.37357 - 0.0889782$i$} & \multicolumn{1}{c|}{\multirow{4}{*}{3}} & \multicolumn{1}{c|}{0}   & \multicolumn{1}{c|}{0.599413 - 0.0926969$i$} & \multicolumn{1}{c|}{\multirow{4}{*}{4}} & \multicolumn{1}{c|}{0}   & 0.809134 - 0.0941578$i$ \\ \cline{2-3} \cline{5-6} \cline{8-9}
\multicolumn{1}{|c|}{}                   & \multicolumn{1}{c|}{1}   & \multicolumn{1}{c|}{0.346071 - 0.273517$i$} & \multicolumn{1}{c|}{}                   & \multicolumn{1}{c|}{1}   & \multicolumn{1}{c|}{0.582598 - 0.281333$i$}  & \multicolumn{1}{c|}{}                   & \multicolumn{1}{c|}{1}   & 0.796583 - 0.284316$i$  \\ \cline{2-3} \cline{5-6} \cline{8-9}
\multicolumn{1}{|c|}{}                   & \multicolumn{1}{c|}{2}   & \multicolumn{1}{c|}{0.298458 - 0.477529$i$} & \multicolumn{1}{c|}{}                   & \multicolumn{1}{c|}{2}   & \multicolumn{1}{c|}{0.551573 - 0.479015$i$}  & \multicolumn{1}{c|}{}                   & \multicolumn{1}{c|}{2}   & 0.772649 - 0.479881$i$  \\ \cline{2-3} \cline{5-6} \cline{8-9}
\multicolumn{1}{|c|}{}                   & \multicolumn{1}{c|}{3}   & \multicolumn{1}{c|}{0.248739 - 0.708946$i$} & \multicolumn{1}{c|}{}                   & \multicolumn{1}{c|}{3}   & \multicolumn{1}{c|}{0.511842 - 0.690648$i$}  & \multicolumn{1}{c|}{}                   & \multicolumn{1}{c|}{3}   & 0.739752 - 0.683955$i$  \\ \hline
\end{tabular}}
\label{table6}
\end{table*}

\vspace{0.5cm}
\section{\label{sec:level7} Bounding the greybody factors and high-energy absorption cross section via Sinc approximation}

\subsection{Bounding the greybody factor}

In this section, the lower bound of the greybody factor of the hairy black hole is investigated. There are many methods to calculate the lower bound, and the WKB method is the most frequently used one \cite{Konoplya:2021ube,Konoplya:2019ppy}. In Ref. \cite{Konoplya:2010kv},  Konoplya et al. studied the greybody factor of the wormhole using the 6th order WKB method.  In Ref. \cite{Dey:2018cws}, they have investigated the greybody factor of Bardeen de Sitter black hole under gravitational perturbation and electromagnetic perturbation using the WKB method. We will use another rigorous method to calculate the greybody. In this method, the general bound of greybody for the black hole is written as \cite{Boonserm:2008zg,Visser:1998ke,Boonserm:2019mon,Ngampitipan:2012dq,Barman:2019vst,Okyay:2021nnh,Javed:2021ymu,Javed:2022bdi,Pantig:2022ely,Javed:2022kzf}
%\begin{equation}
%T \geq \operatorname{sech}^{2}\left(\int_{-\infty}^{\infty} \nu d r_{*}\right),
%\end{equation}
%with
%\begin{equation}
%\nu=\frac{\sqrt{\left(h^{\prime}\left(r_{*}\right)\right)+\left(\omega^{2}-V\left(r_{*}\right)-h^{2}\left(r_{*}\right)\right)^{2}}}{2 h\left(r_{*}\right)}.
%\end{equation}
%Using $h(r_*) = \omega$ with the boundary conditions  $h(\infty) = h(-\infty) = \omega$ yields \cite{}
\begin{equation}
T_{b} \geq \operatorname{sech}^{2}\left(\frac{1}{2 \omega} \int_{-\infty}^{\infty}\left|V\right| \frac{d r}{f(r)} \right).
\end{equation}

Then one can use the Regge-Wheeler potential from section \ref{bridf_review}, and numerically plot the variation of the greybody factor with various parameters ($\alpha$, $l_0$ and $Q$), as seen in Fig. \ref{grey_cross_alpha}, \ref{grey_cross_ll} and \ref{grey_cross_Q}. From these figures, we can see that the value of greybody bound is zero when the frequency is minimal, and the value of greybody bound is 1 when the frequency is large enough. This shows that when the frequency is small, the wave is basically totally reflected. With the increase of frequency, part wave can pass through the potential barrier due to tunneling effect. When the frequency reaches a certain critical value, the wave will not be reflected. In addition, one can see that for larger $\alpha$ and $l_0$, the greybody bound is also larger, whereas the effect of the hair $Q$ is opposite. Therefore, for the hairy black hole spacetime with large charge $Q$, the hairy black hole scatters the incident wave greatly.
On the other hand, the greybody factors for Schwarzschild black holes have been rigorously analyzed \cite{Boonserm:2008zg,Gray:2015pma}. 
Compared with the greybody factors of the Schwarzschild black hole, the greybody factors of the hairy black hole are larger than those of the Schwarzschild black hole when the frequency is fixed, which demonstrates that the probability of Hawking radiation reaching spatial  infinity in the hairy black hole spacetime is greater than that in the Schwarzschild black hole spacetime.

\subsection{The high-energy absorption cross section with the Sinc approximation}

The oscillatory pattern of the high-energy absorption cross section corresponding to a Sinc(x) function within the photon sphere (sinc(x) denoting $ \operatorname{sinc}(x)  \equiv  \operatorname{sin}(x) /x$
). The oscillatory part of the absorption cross section in the eikonal limit is  \cite{Decanini:2011xi}

\begin{equation}
\sigma_{o s c}(\omega)=-8 \pi \sigma_{\text {geo }} n_c e^{-\pi n_c} \operatorname{sinc}\left[2 \pi b_\text{c} \omega\right]
\end{equation}
where
\begin{equation}
n_{c}=\sqrt{f({r_{c}})-\frac{r_{c}^{2}}{2} f^{\prime \prime}({r_{c}})},
\end{equation}
and the eikonal cross-section is $\sigma_{\text {geo }}=\pi b_{c}^{2}$ with the critical impact parameter  $ b_{c}=\frac{r_{c}}{\sqrt{f\left(r_{c}\right)}}$.

\begin{figure*}[htbp!]
\begin{center}
%\hspace{-0.3cm}
\includegraphics[scale=0.65]{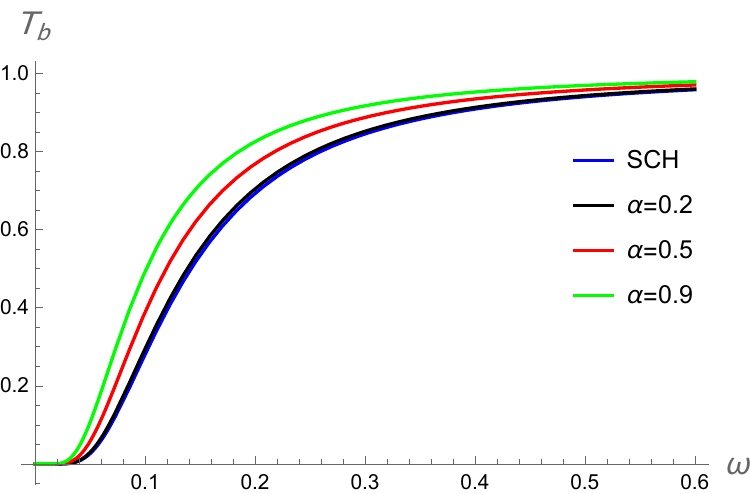}
%\hspace{0.5cm}
\includegraphics[scale=0.65]{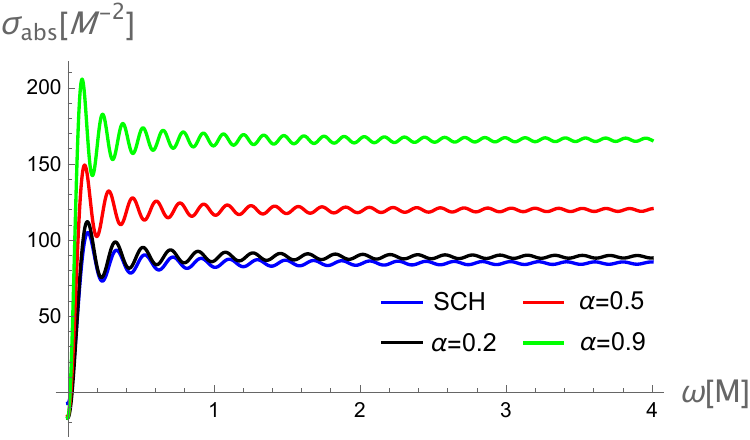}
\end{center}
\setlength{\abovecaptionskip}{0.1cm}
\setlength{\belowcaptionskip}{0.5cm}
\caption{ The greybody bound (left panel) $T_b$ as the function of $\omega$, and the total absorption cross section (right panel) for different values of $\alpha$, with $M = 1, l_0 =1,Q=0.7$.}
\label{grey_cross_alpha}
\end{figure*}

\begin{figure*}[htbp!]
\begin{center}
%\hspace{-0.3cm}
\includegraphics[scale=0.65]{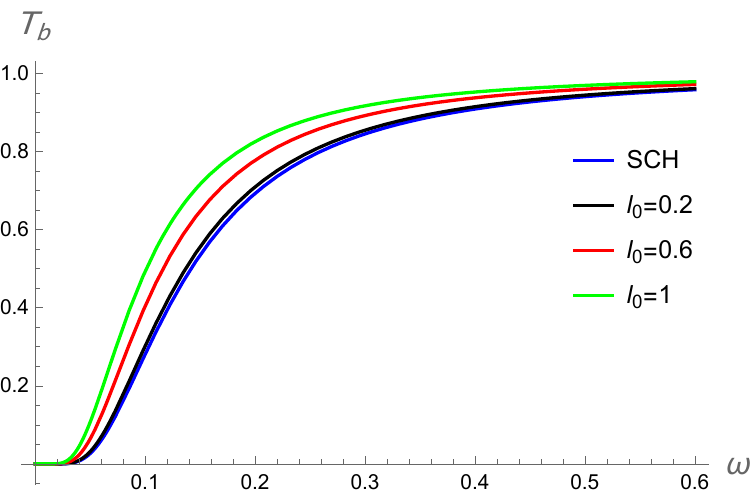}
\includegraphics[scale=0.65]{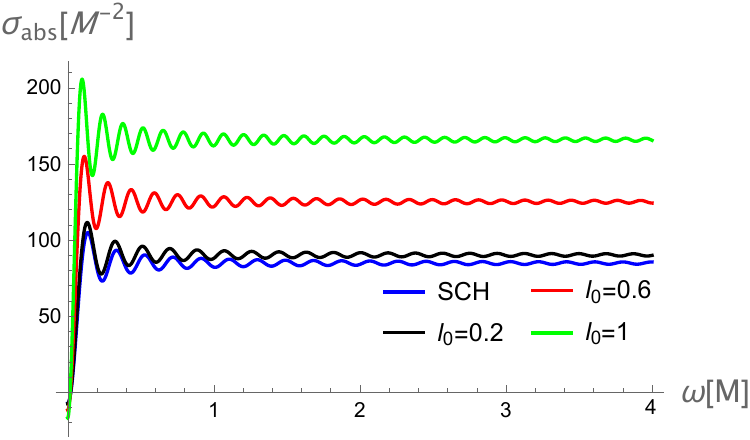}
\end{center}
\setlength{\abovecaptionskip}{0.1cm}
\setlength{\belowcaptionskip}{0.5cm}
\caption{ The greybody bound (left panel) $T_b$ as the function of $\omega$, and the total absorption cross section (right panel) for different values of $l_0$, with $M = 1, \alpha=0.9, Q=0.7$.}
\label{grey_cross_ll}
\end{figure*}

\begin{figure*}[htbp!]
\begin{center}
%\hspace{-0.3cm}
\includegraphics[scale=0.65]{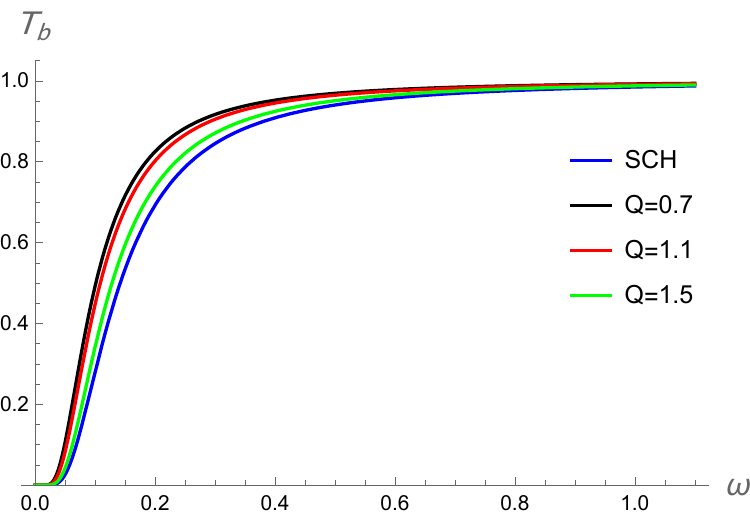}
\includegraphics[scale=0.65]{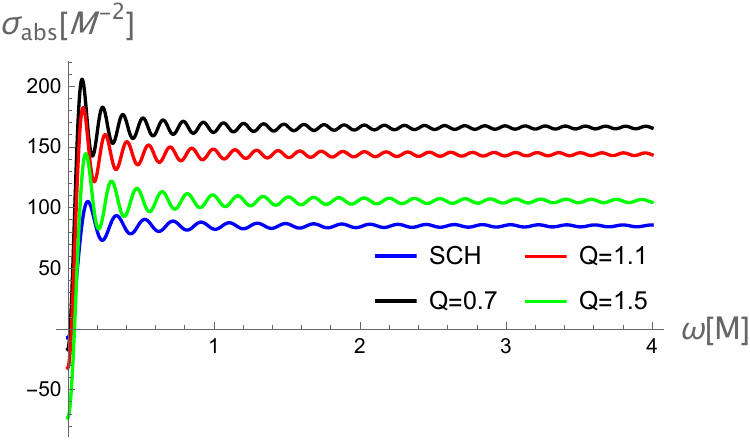}
\end{center}
\setlength{\abovecaptionskip}{0.1cm}
\setlength{\belowcaptionskip}{0.5cm}
\caption{ The greybody bound (left panel) $T_b$ as the function of $\omega$, and the total absorption cross section (right panel) for different values of the charge $Q$, with $ M = 1,\alpha=0.9, l_0 =1$.}
\label{grey_cross_Q}
\end{figure*}

Then the Sinc approximation states that the total absorption cross section at the eikonal limit is $\sigma_{abs} \approx \sigma_{osc} + \sigma_{geo}$ \cite{Decanini:2011xi,Sanchez:1977si,Magalhaes:2020sea,Paula:2020yfr,Lima:2020seq}.  In the Fig. \ref{grey_cross_alpha}, \ref{grey_cross_ll} and \ref{grey_cross_Q} we plot the total absorption cross section for various values of $\alpha$, $l_0$ and $Q$. One can find that the total absorption cross section seems to be divided into three phases with the increase of $\omega$: first, the fast growing phase, then the oscillations phase, and finally the stabilization around a certain value. Moreover, we find that the hairy black hole has a larger absorption cross section when $\alpha$ and $l_0$ are larger, and the absorption cross section of the hairy black hole is smaller when charge $Q$ is larger.
Compared with the total absorption cross section of Schwarzschild black holes, our results show that the total absorption cross section of Schwarzschild black holes is always smaller than that of hairy black holes.

\section{Conclusion} \label{sec:summary}
In summary, we study the QNM of hairy black hole caused by gravitational decoupling. By studying the scalar field, electromagnetic field, and axial gravitational perturbations, the time-domain profiles of QNM are given, and the QNM frequencies of hairy black hole are fitted according to the time-domain profiles, which are consistent with the results of the high-order WKB. We conclude that for hairy black hole caused by gravitational decoupling, the effects of these  hairs ($\alpha,\l_0,Q$) on time-domain profiles and QNM frequencies under scalar field, electromagnetic field, and gravitational perturbations show similar behavior, i.e. an increase in $\alpha$ and $\l_0$ decrease the oscillation frequency of the gravitational wave signal emitted by perturbed hairy black hole, and an increase in $Q$ increase the oscillation frequency of the GW signal.
%Due to the gravitational perturbation with strong gravitational radiation, it is different from the time-domain profiles of the scalar and electromagnetic perturbations lie in the absence of power-law tails.
In addition, the increase of the multipole moment $l$ obviously increases the oscillation frequency of the gravitational wave signal, whereas the influence on its decay rate is very small.
%Our results also show that the hairy black hole spacetime is stable.
%It is worth noting that the time-domain profiles of the hairy black hole remarkably deviate from that of the Schwarzschild black hole under the scalar field perturbation, while the deviation is not as pronounced in the electromagnetic field perturbation and the axial gravitational perturbation as in the scalar field perturbation. Furthermore, we find that the effect of tidal charge $Q$ from the extra sources on the time-domain profiles and the QNM frequencies of the hairy black hole is always opposite to that of charge ($\alpha$ and $l_0$) generating primary hair.
At the end, we study the bounding of greybody factor and absorption cross-section of hairy black hole using the Sinc approximation. It is found that for larger $\alpha$ and $l_0$, the greybody bound and absorption cross section are also larger, whereas the contribution of the charge $Q$ is the opposite. Therefore, for the hairy black hole spacetime with large charge $Q$, the propagating waves can be reflected by the potential barrier greatly. A smaller value of the greybody factor implies that the probability of gravitational radiation reaching spatial infinity is lower. Consequently, an increase in $\alpha$ and $l_0$ can increase the probability of gravitational radiation arriving spatial infinity. In addition, the increase of $\alpha$ and $l_0$ make total absorption cross section also increase, whereas the increase of $Q$ makes total absorption cross section decrease. 
We expect our results to provide some direction for detecting hairy black hole caused by gravitational decoupling in future experiments.
On the other hand, there are some regions of the parameter space in the hairy black hole we study, which will make it become a naked singularity.
In the future, it will be interesting to probe such regions of the parameter space that may pose a threat to the deterministic nature of gravitational theories with spherically-symmetric solutions, by this way one can understand near-extremal modes and strong cosmic censorship in spherical symmetry black hole \cite{Dafermos:2012np,Yang:2013uba,Richartz:2015saa,Zimmerman:2015trm,Hod:2017gvn,Cardoso:2017soq,Cardoso:2018nvb,Dias:2018etb,Destounis:2018qnb}.

%\added{In future, it will be interesting to explore some minimum values of $\alpha$ that turns the naked singularity back into a black hole with an event horizon.}
%-----------------------------------------------------------------------------
%\newpage

%\newpage
%-----------------------------------------------------------------------------

\begin{acknowledgments}
We greatly appreciate anonymous referees for constructive comments.
We also are very grateful to J. Ovalle and Poulami Dutta Roy for useful correspondences. This research was funded by the National Natural Science Foundation of China (Grant No.12265007 and 11565009), the Natural Science Special Research Foundation of Guizhou University (Grant No.X2020068), and Science and Technology Foundation of Guizhou Province (No. ZK[2022]YB029). A. {\"O}. would like to acknowledge the contribution of the COST Action CA18108 - Quantum gravity phenomenology in the multi-messenger approach (QG-MM).

\end{acknowledgments}

%-----------------------------------------------------------------------------

\bibliography{ref}
%\bibliographystyle{apsrev4-1}

%\vspace{7ex}
\end{document}